\documentclass{article}

\usepackage[export]{adjustbox}
\usepackage{multirow}
\usepackage{multicol}
\usepackage{multirow,booktabs,siunitx}

\usepackage{textcomp}

\usepackage{PRIMEarxiv}
\usepackage[hidelinks]{hyperref}
\usepackage{graphicx}       
\usepackage{amsmath}
\usepackage{alltt}
\usepackage{bm}

\usepackage{hyperref}
\usepackage{graphicx}       
\usepackage{amsmath}
\usepackage{alltt}

\usepackage{amsfonts}
\usepackage[svgnames]{xcolor}
\newenvironment{knitrout}{}{}
\usepackage{framed}
\makeatletter
\newenvironment{kframe}{%
 \def\at@end@of@kframe{}%
 \ifinner\ifhmode%
  \def\at@end@of@kframe{\end{minipage}}%
  \begin{minipage}{\columnwidth}%
 \fi\fi%
 \def\FrameCommand##1{\hskip\@totalleftmargin \hskip-\fboxsep
 \colorbox{shadecolor}{##1}\hskip-\fboxsep
     \hskip-\linewidth \hskip-\@totalleftmargin \hskip\columnwidth}%
 \MakeFramed {\advance\hsize-\width
   \@totalleftmargin\z@ \linewidth\hsize
   \@setminipage}}%
 {\par\unskip\endMakeFramed%
 \at@end@of@kframe}
\makeatother

\definecolor{shadecolor}{rgb}{.97, .97, .97}
\definecolor{messagecolor}{rgb}{0, 0, 0}
\definecolor{warningcolor}{rgb}{1, 0, 1}
\definecolor{errorcolor}{rgb}{1, 0, 0}
\makeatletter
\@ifundefined{AddToHook}{}{\AddToHook{package/xcolor/after}{
\definecolor{shadecolor}{rgb}{.97, .97, .97}
\definecolor{messagecolor}{rgb}{0, 0, 0}
\definecolor{warningcolor}{rgb}{1, 0, 1}
\definecolor{errorcolor}{rgb}{1, 0, 0}
}}
\makeatother
\definecolor{fgcolor}{rgb}{0.345, 0.345, 0.345}
\makeatletter
\@ifundefined{AddToHook}{}{\AddToHook{package/xcolor/after}{\definecolor{fgcolor}{rgb}{0.345, 0.345, 0.345}}}
\makeatother
\newcommand{\hlnum}[1]{\textcolor[rgb]{0.686,0.059,0.569}{#1}}%
\newcommand{\hlstr}[1]{\textcolor[rgb]{0.192,0.494,0.8}{#1}}%
\newcommand{\hlcom}[1]{\textcolor[rgb]{0.678,0.584,0.686}{\textit{#1}}}%
\newcommand{\hlopt}[1]{\textcolor[rgb]{0,0,0}{#1}}%
\newcommand{\hlstd}[1]{\textcolor[rgb]{0.345,0.345,0.345}{#1}}%
\newcommand{\hlkwa}[1]{\textcolor[rgb]{0.161,0.373,0.58}{\textbf{#1}}}%
\newcommand{\hlkwb}[1]{\textcolor[rgb]{0.69,0.353,0.396}{#1}}%
\newcommand{\hlkwc}[1]{\textcolor[rgb]{0.333,0.667,0.333}{#1}}%
\newcommand{\hlkwd}[1]{\textcolor[rgb]{0.737,0.353,0.396}{\textbf{#1}}}%

\graphicspath{{media/}}     

\title{Mixture distributions for probabilistic forecasts of disease outbreaks
}

\author{
  Spencer Wadsworth  \\
  Iowa State University \\
  Ames, IA \\
  \texttt{sgw96@iastate.edu} \\
   \And
  Jarad Niemi    \\
  Iowa State University \\
  Ames, IA \\
  \texttt{niemi@iastate.edu} \\
   \And
  Nick Reich \\
  University of Massachusetts Amherst \\
  Amherst, MA\\
  \texttt{nick@umass.edu} \\
}

\begin{document}
\maketitle

\begin{abstract}
Collaboration among multiple teams has played a major role in probabilistic
forecasting events of influenza outbreaks, the COVID-19 pandemic, other disease
outbreaks, and in many other fields. When collecting forecasts from 
individual teams, ensuring that each team's model represents forecast
uncertainty according to the same format allows for direct comparison of 
forecasts as well as methods of constructing multi-model ensemble forecasts.
This paper outlines several common probabilistic forecast representation formats 
including
parametric distributions, sample distributions, bin distributions, 
and  quantiles 
and compares
their use in the context of collaborative projects. We propose the use of
a discrete mixture distribution format in collaborative forecasting in place of
other formats. The flexibility in distribution shape, the ease for scoring and 
building ensemble models, and the reasonably low level of computer storage 
required to store such a forecast make the discrete mixture distribution an 
attractive alternative to the other representation formats.
\end{abstract}




\section{Introduction}


Predicting the outcomes of prospective events is the object of much scientific
inquiry and the basis for many decisions both public and private. Because 
predictions of the future can never be precise, it is usually desirable
that a level of uncertainty be attached to any prediction. In recent years, it
has become increasingly desirable that forecasts be probabilistic in order to 
account for uncertainty in predicted quantities or events 
\cite{gneiting2014probabilistic}. Weather forecasting
\cite{baran2018combining}, economics \cite{groen2013real}, and disease 
outbreaks
\cite{yamana2016superensemble} are some of the areas where probabilistic
forecasting is used.

A probabilistic forecast is a forecast in which possible outcomes are assigned
probabilities. There are a number of ways whereby probabilities or 
uncertainty may be represented. A common representation is either a continuous 
or
discrete parametric distribution, given as a probability density/mass function. 
Much of the literature 
on calibration, sharpness, and scoring of a forecast pertains to parametric 
distribution forecasts
\cite{gneiting2007probabilistic,gneiting2013combining,baran2018combining}.
Other common representations include samples \cite{krueger2016probabilistic}, 
discretized bin distributions \cite{mcgowan2019collaborative},
and quantile forecasts \cite{taylor2021evaluating, bracher2021evaluating}. 
Each representation may be more or less 
appropriate than the others for a given problem, but knowing how to interpret, 
score, and construct ensemble forecasts for a selected representation is 
essential when multiple teams collaborate in the same forecasting project.

Two collaborative projects on forecasting disease outbreaks for which many 
separate forecasts 
are used include the United States Centers for Disease Control (CDC)
annual competition for forecasting the influenza outbreak \cite{cdcflusight}
and the COVID-19 Forecast Hub which has 
continuously operated since the start of the COVID-19 pandemic in the US in 
early 2020 \cite{Cramer2021-hub-dataset}.

\subsection{CDC flu forecasting}

During the 2013-14 flu season, the CDC began hosting an annual competition for 
forecasting the timing, peak, and intensity of the year's flu 
season. The specific events to be forecast were known as \emph{targets}.
Forecasts
for these different targets included forecasts for one, two, three, and 
four weeks into the future. National flu data was provided weekly 
to academic teams not directly affiliated with the CDC who used that data to 
construct forecasts using 
whatever methods they chose. Historically, the forecasts were
submitted 
in a discretized bin distribution or a bin distribution
format. A \emph{bin distribution} is a probability distribution represented by
breaking the numeric range of an outcome into intervals or \emph{bins} and
directly assigning to each bin the probability that the event falls 
within the bin.
During previous flu seasons the \emph{binning scheme} or the assignment of 
bin values was on a numeric 
scale with a bounded range, and the 
prediction of a specific target was a set of probabilities assigned to each bin
\cite{mcgowan2019collaborative}.
These forecasts were then evaluated against actual flu activity, and at 
the end of the season a winning team was declared \cite{cdcflusight}.
The CDC continues to host a flu forecasting project, 
but since the 2021-22 season the 
only 
target for forecasts has weekly hospitalizations and the forecast submission
format has been quantile forecasts similar to those described in the following
section.

Flu forecasting has provided the CDC, competing teams, and other interested
parties a chance to collaborate and improve their forecasting from season to
season. One 
proposed way to enhance prediction has been to aggregate the various teams'
forecasts into a \emph{multi-model ensemble forecast} 
\cite{mcgowan2019collaborative,mcandrew2019adaptively,reich2019accuracy},
or an ensemble forecast.
An \emph{ensemble forecast} is a combination of several component forecast 
models into one model which often yields better predicting power than the 
individual models \cite{cramer2021evaluation}. Such an ensemble made from 
multiple influenza competition forecasts did in fact outperform the individual 
component models \cite{reich2019accuracy}.

\subsection{COVID-19 Forecast Hub}
\begin{figure}[htbp]
\centerline{\includegraphics[scale=.2]{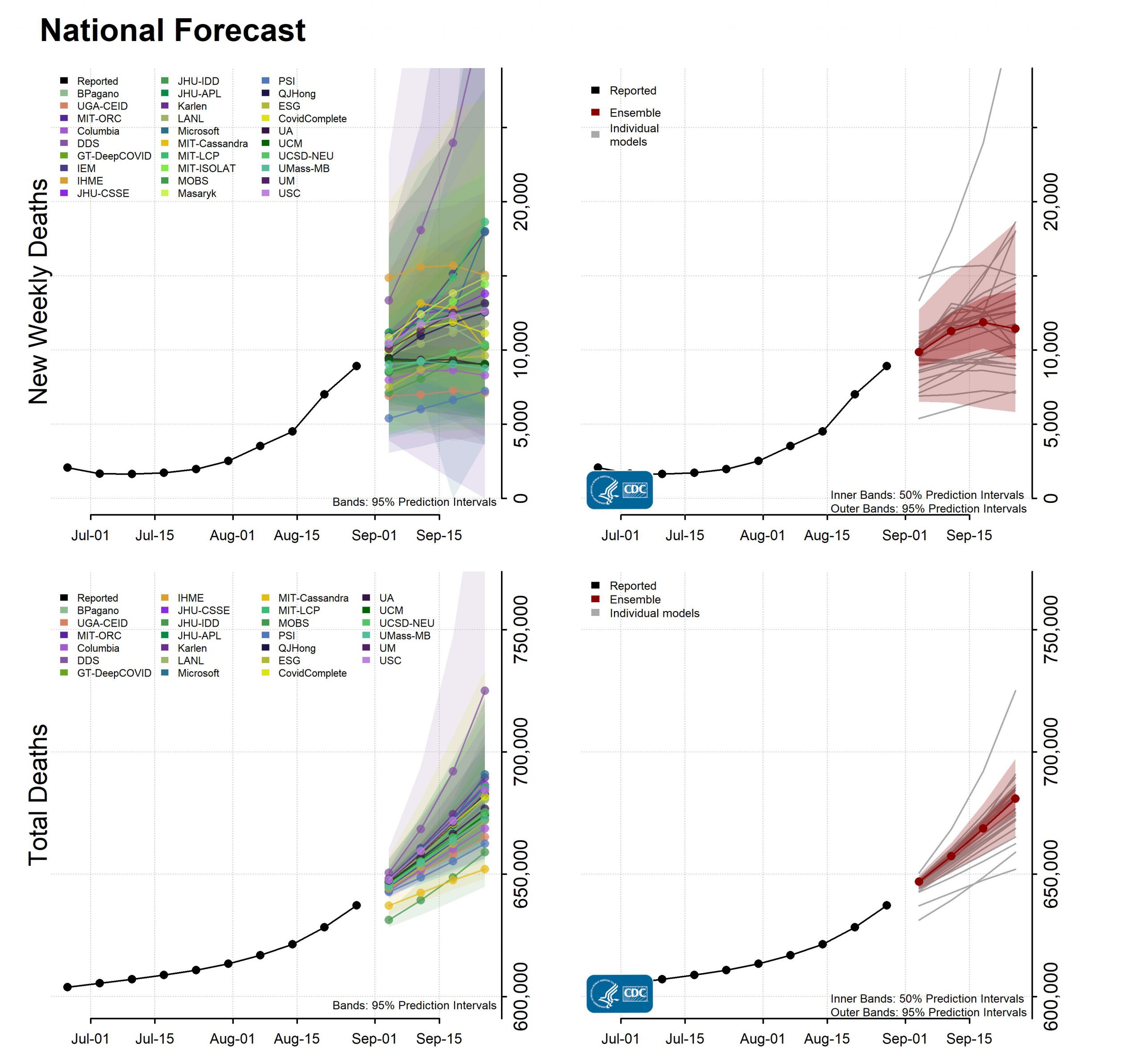}}
\begin{center}
\begin{minipage}{10cm}
\caption[Official CDC COVID-19 deaths report August 2021]{This image,
published on www.cdc.gov in August 2021 as official public communication
by the CDC, shows forecast models for national new weekly deaths due to COVID-19
in the top row and cumulative deaths on the bottom row. The plots in the left
column show prediction intervals from multiple teams whereas the plots on the
right show the intervals for an ensemble forecast model with gray lines that are
individual model point forecast predictions.}
\label{fig:cdcoff}
\end{minipage}
\end{center}
\end{figure}

In March 2020, at the onset of the COVID-19 pandemic, the United States
COVID-19 Forecast Hub
was founded. Borrowing from the work done in the CDC flu competition, the 
COVID-19 Forecast Hub was a central site in 
which dozens of academic teams collaborated to forecast the ongoing COVID-19 
pandemic.
Every week relevant
pandemic data aws provided to these teams who constructed forecast models to 
predict the target cases, hospitalizations, and deaths due to COVID-19. 
Forecasts were
made on the US county,
state, and national levels and for days, weeks, and months ahead.
These forecasts were aggregated into a single ensemble forecast. The model data,
forecasts, and the ensemble forecast were passed along to the CDC for its use in 
official
communication \cite{Cramer2021-hub-dataset}. Figure \ref{fig:cdcoff} is from 
an 
official CDC report from August 2021. It shows forecasts from the COVID-19 
Forecast Hub of increment deaths and cumulative deaths due to COVID-19.

Though similar to the forecasting in the CDC flu competition, the format of 
the COVID-19 Forecast Hub has key distinctions. First, this project has
been operating continuously since it began, so forecasts have been made
for over 100 straight weeks. 
Second, rather than bin distributions
the forecasts are requested as the predictive median and 
predictive intervals for various nominal levels depending on the target to be
predicted \cite{bracher2021evaluating}. Each value in a predictive interval
is a value for a quantile at a specified nominal level. This makes a set of 
predictive intervals a \emph{quantile forecast} or a forecast made up of a set
of quantiles and corresponding values.
Collecting forecasts as quantile forecasts instead of bin forecasts brings with 
it differences in how to score the forecasts, construct an 
ensemble forecast, and store the forecasts among other differences.
Ray et. al show that ensemble forecasts in the COVID-19 Forecast Hub provide
precise short-term forecasts which decline in accuracy in longer term forecasts
approaching four weeks \cite{ray2020ensemble}.

\subsection{Outline}
In the context of collaborative forecasting like that of the CDC flu competition
or the COVID-19 Forecast Hub, bin forecasts and quantile forecasts have become
important representations. Yet both 
representations come 
with their drawbacks. Computer
storage for instance might be a concern if many bin distributions are used for 
forecasting, 
and scoring methods are limited if forecasts are quantile forecasts. 
In this paper, we propose the use of finite mixture distributions as a means 
of forecasting for collaborative projects similar to the CDC flu competition or 
the COVID-19
Forecast Hub. A finite mixture distribution --which we will refer to as a
\emph{mixture distribution}-- is a distribution constructed by aggregating a 
finite 
collection of other distributions. In this paper, we focus on the case where the
collection of distributions are parametric distributions.
In Section \ref{section:representations}, 
popular probabilistic forecast representations 
are defined and reviewed. For each representation, we review methods 
of scoring, storing, constructing 
ensembles, and other aspects.
Section \ref{section:conmixforc} presents using mixture 
distributions in a collaborative forecast project and discusses tools
 for scoring and constructing an ensemble forecast.
Section \ref{section:retrostud} is a retrospective study of the CDC flu 
competition and COVID-19 Forecast Hub
forecasts and an attempt to assess whether forecast models
may be approximated by one component mixture distributions.


\section{Probabilistic forecast representations}
\label{section:representations}

At least four forecast representations are commonplace in 
forecasting. In a collaborative setting, certain aspects of each representation
should be considered including scoring, computer storage, and how to construct
an ensemble forecast. For each representation discussed herein, applications of 
each of those aspects are also discussed.

\subsection{Things to consider in collaborative projects}

In this section we introduce three topics which ought to be considered in a 
collaborative forecast project including forecast scoring, computer storage of
forecasts, and ensemble forecast construction from multiple models. 

\subsubsection{Scoring}
Scoring rules are used to numerically evaluate or \emph{score} a probability 
forecast. 
The
score is a measure of the accuracy of the forecast and where multiple forecasts
exist the score for each may be used to compare forecasts. If a scoring rule 
is \emph{proper}, then the best possible score is obtained by reporting the true 
distribution. The rule is strictly proper if the best score is unique. 
Under proper
scoring rules, a forecaster has no incentive to be dishonest in their 
submission \cite{gneiting2007strictly}. 
This makes proper scoring rules ideal
for evaluating forecasts. We will limit our review of scoring methods to rules
which are proper.

\subsubsection{Storage}
For a collaborative forecast project where many researchers are involved and 
many predictions are collected, computer storage may need to be addressed. 
As an example of required computer storage, the 
repository for the
COVID-19 Forecast Hub contained 85 million forecasts as of 
April 4, 2022 which required more than 11.7 gigabytes of storage.
\cite{covidgithub}.
When determining the goals of a forecast project, there should be consideration 
of the storage required for different forecast representations.

\subsubsection{Ensemble model}
An ensemble model is a statistical model made by combining information from
two or more individual statistical models. Private and public decisions are 
regularly made after combining information from multiple sources. For a given
problem, information from one source may provide insight on a subject which 
other
sources fail to capture. Likewise one statistical model may provide insight
that another model does not. Thus when multiple models are combined, the 
resulting ensemble may outperform the individual models.

As probabilistic forecasting becomes more commonplace, so too does ensemble 
modeling. Ensembles have been used extensively in weather and
climate modeling \cite{baran2018combining}, 
and they have been used increasingly in modeling infectious disease outbreaks
\cite{yamana2016superensemble}. 
Ensembles allow for an incorporation of multiple signals --often from differing
data sources-- and sometimes individual model biases are 
canceled
out or reduced by biases from other models \cite[see references
therein]{reich2019accuracy}. 
In several disease outbreak studies, ensemble forecasts have been 
shown to outperform individual model forecasts
\cite[see
references therein]{ray2020ensemble, cramer2021evaluation}.
Construction of an ensemble may be done by combining individual forecast 
models using weighted averages. This has been called stacking 
\cite{wolpert1992stacked} or weighted density ensembles 
\cite{ray2018prediction}. 

Considered the 
state-of-the-art techniques for combining 
component distributions
into an ensemble distribution are nonhomogeneous regression and ensemble
model averaging (MA), both of which are defined by Gneiting and Katzfuss
\cite{gneiting2014probabilistic}.
In the context of an ensemble made from component models submitted from various
sources,
MA may be preferable because it does not require that modeling methods for 
individual components be the same. The general form for an MA ensemble distribution 
$p^E$
is defined in equation \eqref{eq:bma}, where $p_m$ is the $m^{th}$ component 
forecast distribution and 
$0 \leq w_m \leq 1$ is a weight assigned to that component where $\sum_m w_m = 1$.

\begin{equation}
\label{eq:bma}
  p^E(x) = \sum_{m=1}^M w_mp_m(x)
\end{equation}

In MA, the final model does not have to be specified
beforehand and the resulting forecast will be a mixture distribution
of all component forecasts. 
Many methods exist for estimating weights. Some of these methods 
include maximum likelihood estimation
\cite{raftery2005using}, Markov chain Monte Carlo (MCMC) 
sampling \cite{vrugt2008ensemble}, Bayesian model averaging, Akaike or AIC
weights, \cite{wagenmakers2004aic}, minimizing the CRPS of the ensemble
\cite{baran2018combining}, and others.

\subsection{Probabilistic forecast representations}


Probabilistic forecast uncertainty can take on many representations. In this
section we describe parametric distributions, sample distributions, bin 
distributions, and predictive intervals as representation forecasts. Following 
the description of these four representations, mixture distributions are 
introduced.

\subsubsection{Parametric distributions}
\label{section:pardist}
A parametric distribution is a discrete or continuous probability distribution 
described by a known function $p(x) := p(x|\theta)$. The function $p(x)$ is 
called a probability mass function (pmf) if the distribution is discrete and 
a probability density function
(pdf) if the distribution is continuous. Here $\theta$ is a vector of 
parameters contained in the parameter space of the distribution. 

%
%
%
%
%
%
%
%
%

For a forecast represented as a parametric distribution 
with pmf/pdf $p_m(\cdot)$, 
the accuracy of the forecast may be measured by how likely the 
realized value $x^*$
is to occur. Commonly used proper scoring rules for parametric distributions
include the logarithmic score (LogS), the
continuous rank probability score (CRPS) \cite{hersbach2000decomposition}
\cite{alves2013ncep}, and the interval/Brier 
score (IS) \cite{gneiting2007strictly} among 
others. See also \cite{gneiting2014probabilistic}
Section 3 for more on proper scoring functions. The definitions 
\eqref{eq:logs}, \eqref{eq:crps}, and \eqref{eq:is} are found in the review by 
Krueger.
For a forecast with pdf/pmf $p(\cdot)$, \eqref{eq:logs} evaluates the 
probability of the observed value $x^*$.

\begin{equation}
\label{eq:logs}
  \text{LogS}(p,x^*) = -\log p(x^*)
\end{equation}

The goal for a forecaster is to minimize the LogS, so a forecast $p'(x^*)$ 
is considered 
superior to $p(x^*)$ if
$\text{LogS}(p', x^*) < \text{LogS}(p, x^*)$.
The LogS is limited to scoring forecasts with density functions and
evaluating those densities only at the point $x^*$.
For the cumulative distribution (CDF) $F$ of a parametric distribution, 
the CRPS is defined in \eqref{eq:crps}.
Here too a smaller score indicates a more accurate forecast.

\begin{equation}
\label{eq:crps}
  \text{CRPS}(F, x^*) =\int_{-\infty}^{\infty} (F(x)- 1_{\{x*\leq x\}})^2 dx
\end{equation}

Standard practice for constructing an ensemble model from multiple parametric
models is to use MA.
For selecting distribution weights for the ensemble, minimizing the LogS or 
CRPS of 
$p^E(x)$ is common. 
Because it is evaluated over the whole distribution, 
minimizing the CRPS has some nice properties, but it can also be
difficult to compute. For example, when the forecast is a mixture of a 
truncated normal distribution (TN) and a truncated lognormal (TL), the CRPS is 
not available in closed form \cite{baran2018combining}.

Generally computation and evaluation of parametric distribution functions is easy. 
For most commonly used parametric distributions --normal, lognormal, Poisson,
gamma, etc.-- there is software readily available to compute density, 
distribution, and quantile values. 
A completely defined continuous parametric distribution may be evaluated at a
continuously
infinite number of values which we call an \emph{infinite resolution}.
Requirements for storage are also low 
compared to other representations that will be discussed since the most 
common
parametric distributions can be fully defined with three or four pieces of 
information including the distribution family and the corresponding parameters. 
Table \ref{table:pstor} contains enough information to completely define a 
$\mbox{Lognormal(1,0.4)}$ 
truncated to the inteval $[0,8]$. 
The truncation is done here so as to make a
direct comparison with the distributions shown later in Tables 
\ref{table:dbins} and \ref{table:qstor}.

\begin{table}[h!]
\begin{center}
\begin{minipage}{10cm}
\centering
\caption[Parametric distribution storage]{This is an example of information required
 to specify a truncated lognormal distribution.}
 \begin{tabular}{|c c c c c|} 
 \hline
 family & param1 & param2 & lowerlim & upperlim\\ [0.5ex] 
 \hline
 lnorm & 1 & 0.4 & 0 & 8 \\ 
 \hline
 \end{tabular}
 
 \label{table:pstor}
 \end{minipage}
 \end{center}
\end{table}

A drawback of representing a forecast in a parametric distribution is the lack 
of flexibility in the model selection. Easy computation and evaluation of these 
models is limited to what is available in software, so certain distributional
shapes may be unattainable.
Requiring a parametric forecast also bars the use of some statistical methods
which might be used to create a forecast model including some Bayesian methods
where a posterior distribution cannot be computed in closed form.

\subsubsection{Sample distributions}
Forecasters may want more flexibility in modeling than a 
parametric 
distribution can provide. Methods that require sampling from a posterior 
distribution or bootstrap sampling to obtain a forecast distribution are examples
where parametric distributions may not be appropriate for modeling because of 
the lack of flexibility in distribution shape.

A sample distribution is composed of a collection of,
possibly weighted, 
random variables 
$(X_1,...,X_n)$ where $X_i \sim D$ and $D$ is some distribution. From this 
sample,
statistics such as mean, median, variance, and quantiles may be calculated. 
An empirical cumulative distribution function (ECDF) may also be calculated as
in \eqref{eq:ecdf}. If a sufficiently large 
sample is generated from a distribution for which an expectation exists, 
the sample will closely approximate the 
true distribution.

\begin{equation}
\label{eq:ecdf}
  \text{ECDF} = F_n(x) = \frac{1}{n} \sum_{i=1}^n \mathbb{I}(X_i \leq x)
\end{equation}


For common distribution families it is easy to generate large samples using 
existing functions in \texttt{R} and other programming platforms. For some 
distributions 
for which the mathematical formula is unknown or is not in closed form, more 
sophisticated methods may be required to generate samples. Bayesian analyses may 
require MCMC sampling to generate a 
sample. Such samples are useful in that the true distribution may be closely 
approximated without knowing the true mathematical form. 

Under the sample distribution representation, the options that a researcher has 
for constructing a forecast are more than
if they are asked to submit a parametric distribution, and the range of possible 
shapes for a distribution is larger.
In the last few decades, increased computing power and improvements in MCMC
sampling have greatly contributed to growth in the use of
sample distributions for forecasting \cite{gneiting2007strictly}
\cite[see examples listed therein]{krueger2016probabilistic}.

To properly score a forecast represented by a sample distribution, both the CRPS
and LogS may be used. The CRPS has the advantage here of scoring the sample 
distribution directly since the CDF in  (\ref{eq:crps}) may be replaced with the 
ECDF in
(\ref{eq:ecdf}). To use the LogS to score a forecast, a density function for the 
sample may be approximated. Common approximations include a kernel density (KD)
or Gaussian approximation (GA) 
\cite[for example]{krueger2016probabilistic}.

The KD in \eqref{eq:kd} is defined by Krueger et. al 
where $K$ is a kernel function, and $h_n$ is a suitable bandwidth. The GA is 
defined in \eqref{eq:ga} where $\Phi$ is the standard normal CDF and 
$\hat{\mu}_n$ and $\hat{\sigma}_n$
are the empirical mean and standard deviation of the sample $(X_i)$
\cite[see also for a comparison of scoring MCMC drawn forecasts between the CRPS 
and the LogS]{krueger2016probabilistic}.

\begin{equation}
\label{eq:kd}
  \hat{p}_n^{KD}(x) = \frac{1}{n h_n} \sum_{i=1}^n K \left( \frac{x-X_i}{h_n} \right) 
\end{equation}

\begin{equation}
\label{eq:ga}
  \hat{F}_n^{GA}(x) = \Phi  \left( \frac{x-\hat{\mu}_n}{\hat{\sigma}_n} \right)
\end{equation}

To build an ensemble model from sample distribution forecasts, the MA 
construction from (\ref{eq:bma}) may be used only replacing $p_m$ with the
approximate KD or GA pdf functions
-$\hat{p}_{n_m}^{KD}$ or $\hat{p}_{n_m}^{GA}$ respectively where 
$\hat{p}_n^{GA}$ is the pdf corresponding to (\ref{eq:ga}).  
The optimal weights 
$w_m$ may be estimated by maximizing the likelihood or minimizing the CRPS. 
If the desire is that the ensemble has uniform weights, 
a random sample from $(X_n)_m$ with probability $w_m$ can be obtained.

A potentially large issue with using sample distributions is the amount of 
storage
it may require. For example when making MCMC draws from a posterior 
distribution, the final sample distribution can have a sample size of thousands
or tens of thousands. Maybe not all distributions would require such a large 
sample size, but sizes of at least dozens or hundreds would be required for each
forecast prediction. For any project the size of the CDC flu competition or the
COVID-19 Forecast Hub, the storage required would be large and potentially 
expensive.

\subsubsection{Bin distributions}
An alternative to parametric distributions and sample distributions,
which allows
for higher flexibility in distribution shape than a parametric distribution and 
will usually require less storage space than samples is the bin 
distribution.
A bin distribution may be constructed over a set 
$A = [a, b)$ by partitioning $A$ into a set of $K$ bins $\{B_i\}_{i=1}^{K}$
where $B_i = [b_{i-1}, b_i)$ and $\cup_{i=1}^{K} B_i = A$. Based on the problem
to be forecast, researchers will determine the possible range $A$ and 
select the number of bins and the sizes for each bin. It may be the case
that a collaborative project will set the widths of all bins to be equal so that 
$\Delta = b_i - b_{i-1}$ is the same width for all $i$ 
\cite{mcgowan2019collaborative}.
To complete the construction, a probability $p_i$ is assigned to each $B_i$ 
where $\sum p_i = 1$. These probabilities are determined by the 
forecasters. 
This representation with a given bin and assigned probability may be 
treated like a discrete distribution with a pmf in that the calculation of 
the cumulative distribution is done similarly to that of a discrete parametric 
distribution. The 
cumulative distribution may be calculated as in \eqref{eq:bcdf}.
Here $p_i$ is the probability for the bin $B_i$ where $x \in B_i$.

\begin{equation}
\label{eq:bcdf}
  P(X \leq x) = \sum_{i =1}^{n:x \in B_n} p_i
\end{equation}

If the value to be forecast takes on discrete values, a common discrete 
distribution, such as a Bernoulli or Poisson distribution, may sometimes be used 
to assign probabilities to each of the bins. When the values to be forecast are
continuous, a forecaster may need to employ a method of discretization to a 
forecast distribution. There are a number of possible ways to do this including
those outlined by Chakraborty and Subrata \cite{chakraborty2015generating}.

For the first several seasons when the CDC hosted the flu competition, the 
forecast representation used was the bin distribution. The CDC has also used the
bin distribution represetation for other disease outbreak
forecast projects.
In that context it has become the standard representation 
\cite{brooks2020comparing}. Much work has been done in evaluating 
and constructing ensemble models on influenza forecasts represented by 
discretized bins 
\cite{mcgowan2019collaborative, mcandrew2019adaptively,reich2019collaborative}. 

Because a bin distribution can be viewed as a pmf, methods for proper scoring 
already 
discussed --LogS and CRPS-- are useable and MA is a valid method for ensemble 
construction. Reich et. al used MA to combine multiple forecasts from the flu 
competition. They constructed and compared ensemble models with different 
weighting
schemes including equally weighted components, $w_m = 1/M$,
and estimating weights
according to the model specification. To estimate weights they used the 
expectation maximization (EM) algorithm 
\cite[see
supplementary material within for details]{reich2019accuracy}.

The exact amount of information required for a bin forecast will 
vary depending on the permitted range of the forecast and the desired 
resolution. 
In the CDC flu contest, a forecast might have 131 bins between 0\% and 13\% 
--bins having increments of 0.1 or 0.01\%-- with 
corresponding probabilities in each. This makes 262 pieces of information per
prediction. For any binning scheme of more than two or three bins, the 
information 
requirement for a bin distribution will be higher than for a parametric 
distribution.
Table \ref{table:dbins} illustrates what the bin distribution discretized 
from a 
$\mbox{Lognormal(1,0.4)}$ distribution truncated over $[0,8]$
looks like in 41 equally spaced bins. The discretization was done
such that the probabilities $p_i$ are calculated as in \eqref{eq:disc}
where $p^{TL}$ is the pdf of a truncated $\mbox{Lognormal(1,0.4)}$. This is 
similar
to Methodology-IV from Chakraborty 
\cite{chakraborty2015generating,kemp2004classes}. The truncation here is done
because in practice a bin forecast will generally have a finite support.

\begin{equation}
\label{eq:disc}
  p_i = \int_{b_{i-1}}^{b_i} p^{TL}(x) dx
\end{equation}

Submitted as a forecast prediction, the distribution illustrated in Table 
\ref{table:dbins} 
includes 82 values. For parts 
of the CDC influenza competition some forecasts included up to 262. This is 
far less storage than the possible thousands of draws from a sample 
distribution but is 
still much larger than the three or five pieces of information required to 
report a 
lognormal
or truncated lognormal distribution.

\begin{table}[h!]
\begin{center}
\begin{minipage}{10cm}
\centering
\caption[Bin distribution storage]{This is a storage example of a
  discretized
  lognormal with $\mu = 1$ and $\sigma = 0.4$ truncated over $[0,8]$.}
 \begin{tabular}{|c|c|} 
 \hline
    bin & prob \\ \hline
    ... & ... \\
    {[1.4,\;1.6)} & 0.04414 \\
    {[1.6,\;1.8)} & 0.05896 \\
    {[1.8,\;2.0)} & 0.07032 \\
    {[2.0,\;2.2)} & 0.07172 \\
    {[2.2,\;2.4)} & 0.07955 \\
    ... & ... \\
 \hline

 \end{tabular}
  
\label{table:dbins}
\end{minipage}
\end{center}
\end{table}

Besides the potentially large amount of information required per forecast,
creating the right 
binning scheme may be a challenge. Because there must be a finite number of
bins, forecast distributions often have finite support. And where the range of 
possible outcomes to a problem is not well known, the right binning scheme may 
be
hard to produce. This may depend on the details of the event to be forecast, 
but in the case of the COVID-19 outbreak, choosing the right set of bins posed
a few problems.

\subsubsection{Interval forecasts}
When deciding how forecasts should be represented in the COVID-19 Forecast Hub,
the time pressure of generating forecasts and the large
range for possible outcomes both contributed to the COVID-19 Forecast Hub 
decision
to forego trying to create the right binning scheme and use quantile forecasts
to forecast the COVID-19 pandemic \cite{bracher2021evaluating}.
The COVID-19 Forecast Hub requires predictions to be submitted as 11 or three 
nominal 
intervals --depending on the specific target and unit to be forecast-- 
and a median. Likewise the CDC flu forecasting project has used this same 
format for the past two seasons \cite{cdcepiflu}.

A quantile forecast is constructed as in \eqref{eq:quantiles}.
Here for $N$ given quantile levels $\alpha_1,..., \alpha_N$; $q_1,..., q_N$ 
are the 
values such that we have \eqref{eq:quantiles}. 
When the quantiles are reported as prediction intervals we have 
\eqref{eq:intervals}.

\begin{equation}
\label{eq:quantiles}
  P(Y \leq q_1) = \alpha_1, P(Y \leq q_2) = \alpha_2, ..., 
  P(Y \leq q_N) = \alpha_N
\end{equation}

\begin{equation}
\label{eq:intervals}
  P(Y \leq q_1) = \alpha_1, P(Y \leq q_2) = \alpha_2, ...,
  P(Y \leq q_{N-1}) = 1 - \alpha_2, P(Y \leq q_N) = 1 - \alpha_1
\end{equation}

The CRPS and LogS may not be used to score a quantile forecast, and MA may not
be used for constructing an ensemble from multiple quantile forecasts.
Methods for scoring quantile forecasts and constructing ensemble
forecasts are limited, and when given only in quantiles 
the shape of a distribution is not known. In fact
nothing is known about the tails or the uncertainty beyond the most 
extreme reported quantile values. In the COVID-19 Forecast Hub forecasts, 
nothing 
is reported about the range below the $1^{st}$ quantile or above the $99^{th}$.
Yet the quantile representation has its advantages. 
Quantile forecasts allow for forecasters to submit fairly detailed
forecasts without restricting the range of possible values.
Since quantiles are easily calculated from any regular distribution type
--using the quantile function for parametric functions or calculating sample 
quantiles-- we consider quantile forecasts to be highly flexible in terms of 
what methods forecasters may employ in modeling.

To score a quantile forecast, 
neither the LogS nor the CRPS may be used, 
but another proper scoring rule the IS may be used.
For an observed outcome $x^*$ and a prediction interval $(l,r)$ 
where $l$ and $r$ are the $\alpha/2$ and $(1-\alpha/2)$ quantiles that bound
the central $(1-\alpha)$ prediction interval, the IS is defined as 
in \eqref{eq:is}.
This is a sum --weighted by 
$\alpha$-- of the width of the
interval and the distance between $x^*$ and the interval (if $x^*$ is not 
captured in the interval) \cite{gneiting2014probabilistic}. 
The IS requires only a single central 
$(1-\alpha) \times 100$ prediction interval.

\begin{equation}
\label{eq:is}
  IS_{\alpha}(l,r; x^*) = (r-l) + \frac{2}{\alpha}(l-x^*)1\{x^*<l\} 
  + \frac{2}{\alpha}(x^*-r)1\{x^* > r\}
\end{equation}

When a quantile forecast is made up of multiple intervals each with different
$\alpha$ levels, the weighted interval score (WIS) may be used.
Bracher et.
al use the WIS to score COVID-19 quantile forecasts 
\cite{bracher2021evaluating}.
There are multiple versions of the WIS, some of 
which are described in Bracher et. al, but the version used by the COVID-19 
Forecast
Hub for a forecast of $K$ intervals is defined in \eqref{eq:wis}.
Here $median$ refers to the predictive median and $w_k = \alpha_k/2$ is the 
weight on the $k^{th}$
interval. With that selection of weights, it may be shown that the
WIS approximates the CRPS \cite[see S1 Text therein]{bracher2021evaluating}.
\begin{equation}
\label{eq:wis}
  WIS_{0,K}(F_m,x^*) = \frac{1}{K + 1/2} \left(w_0 |x^*-median|+
  \sum_{k=1}^K \{ w_k IS_{\alpha_k}(F_m,x^*) \} \right)
\end{equation}

Bogner, Liechti, and Zappa compared scoring forecasts of quantiles with the 
Quantile Score (QS) similar to the interval score and scoring distribution functions
fit to those quantiles using the CRPS \cite{bogner2017combining}. The CRPS 
corresponds to the integral of the QS over all possible thresholds rather than
just specific quantiles, so it more effectively reveals deficiencies in parts of 
the distribution and especially in the tails past the end points of quantiles
used in QS or IS. Thus there may be something lost in terms of scoring when the 
WIS is used since it also is constructed from the IS.

Like the CRPS, not only does the WIS provide an easily interpretable proper 
score for interval forecasts, but it may also be useful when building an 
ensemble forecast.
The ensemble forecast constructed by the COVID-19 Forecast Hub is made as an
equally-weighted average of forecasts from the component models. More 
specifically, each quantile value of the ensemble is the average of values from
all models corresponding to the same quantile \cite{ray2020ensemble}. For $M$ 
models each with $K$ quantiles, the $k^{th}$ ensemble quantile $q^E_k$ is 
calculated as in \eqref{eq:qa}
where $w_m$ is the weight assigned to each forecast and $\sum w_m = 1$. In
the COVID-19 Forecast Hub model, $w_m = w = 1/M$. Where the overall mean or 
a weighted mean may be used for averaging, the median may also be used.
Brooks et. al compare performance of the COVID-19 ensemble
using equally-weighted means, weighted means, and median value constructions
\cite{brooks2020comparing}.
In their report, they show that weighted means and median constructions tend
to outperform an equally-weighted mean construction. To come up with optimal 
weights, they select values $w_m$ from (\ref{eq:qa}) which minimize the WIS of 
the ensemble forecast.

\begin{equation}
\label{eq:qa}
  q^E_k = \sum_{m=1}^M w_m q_k^m 
\end{equation}

%

As in sample distributions and bin distributions, 
data storage for interval forecasts will
depend on the desired clarity of resolution. For the COVID-19 forecasts 
submitted to the COVID-19 Forecast Hub, 23 quantile values are requested for 
quantiles (0.01, 0.025, 0.05, 0.10, …, 0.95, 0.975, 0.99). This includes a 
median along with 11 predictive intervals \cite{bracher2021evaluating}. 
Forecasters are thus required to submit 46 values in each short-term forecast 
(some of the longer term forecasts only include seven quantiles). In terms of 
storage, this is an improvement over requirements for the CDC flu competition. 
Table
\ref{table:qstor} shows how a submission of 23 quantiles from a 
$\mbox{Lognormal(1,0.4)}$ truncated on $[0,8]$ might look.

\begin{table}[h!]
\centering
\begin{center}

\caption[Quantile storage]{This shows six quantiles
 and values from a lognormal distribution with $\mu = 1$ and $\sigma = 0.4$}
 \begin{tabular}{|c||c|c|c|c|c|c|c|c|c|}
 \hline
    quantile & 0.01 & 0.025 & 0.05  & ...  & 0.95 & 0.975 & 0.99 \\ \hline
    value & 1.07137 & 1.2404 & 1.40689 & ... & 5.18328 &
    5.82391 & 6.58783 \\
    
 \hline
 \end{tabular}
 \begin{minipage}{10cm}
 
 \label{table:qstor}
 \end{minipage}
 \end{center}
\end{table}

\begin{figure}[htbp]
\begin{center}

\centerline{\includegraphics[scale=.2]{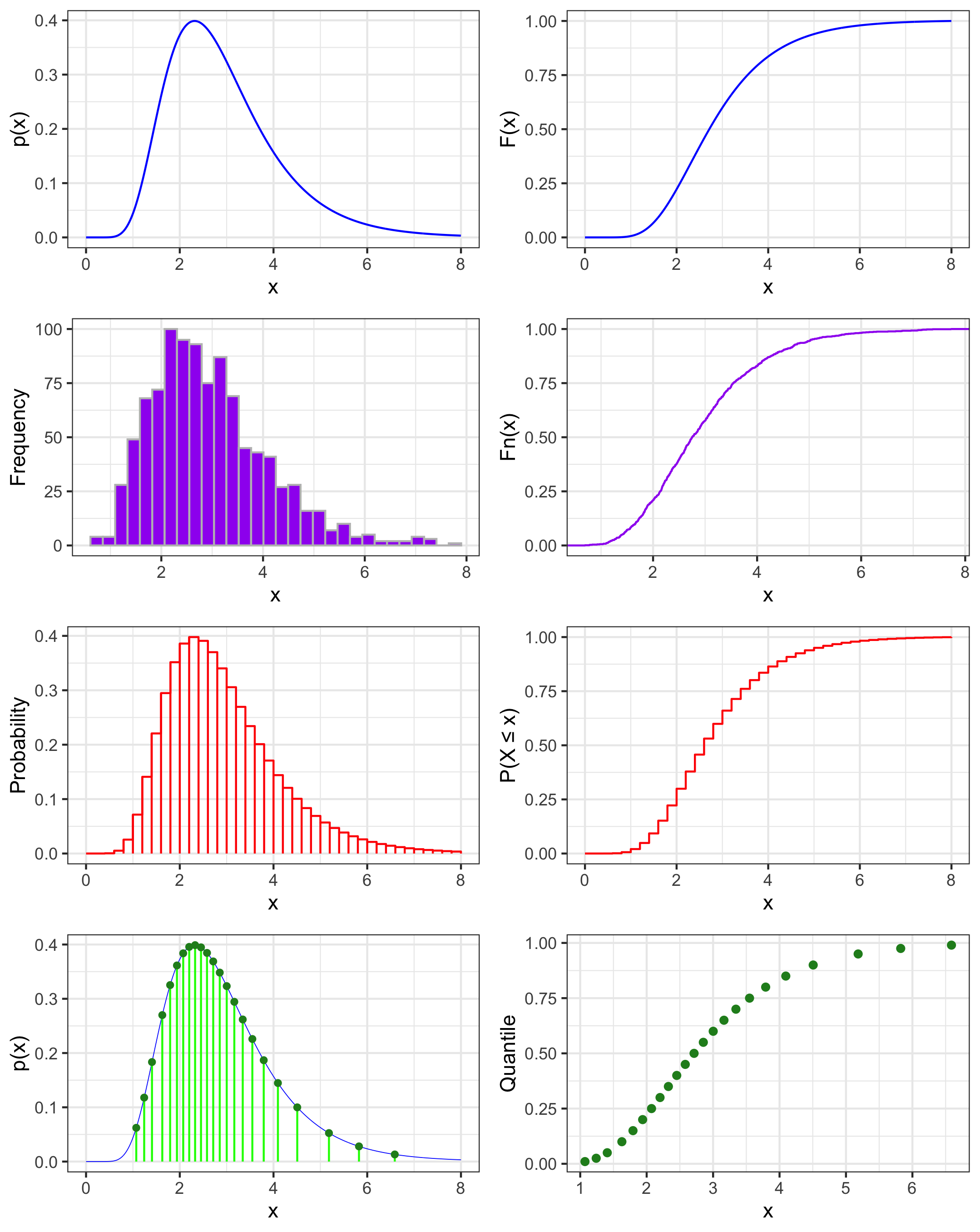}}
\begin{minipage}{10cm}
\caption[Density/CDF comparison between parametric distribution, sample 
distribution, discretized bin distribution and quantiles]{This figure compares 
the
densities and CDFs of forecast representations discussed in the left and 
right columns respectively. Each is generated from a Lognorma(1,0.4) 
distribution truncated on $[0,8]$. 
Blue shows the density and CDF functions. Purple shows a 
histogram and ECDF of 1,000 samples. Red shows bin probabilities and the CDF
function for a bin distribution. Green shows quantiles
with corresponding values.}
\label{fig:denscomp}
\end{minipage}
\end{center}
\end{figure}

Figure \ref{fig:denscomp} illustrates how the densities and CDFs compare 
between parametric distributions, sample distributions, bin distributions, and
quantiles. 

\newpage

\subsection{Mixture distributions}
A mixture distribution forecast representation is an
attractive alternative to the four representations already discussed. 
A mixture distribution forecast
would allow for a large range of distribution shapes, a high resolution, 
storage comparable
to that of bin and quantile forecasts, and ensemble construction using MA.
A mixture distribution may be constructed in the same way as the
ensemble described in section \ref{section:pardist} 
(\ref{eq:bma}) 
where for $C$ distributions
with pdfs $p_c(x)$ and $w_c > 0$ and $\sum w_c = 1$ we have 
\eqref{eq:dmd}.

\begin{equation}
\label{eq:dmd}
  p^{M}(x) = \sum_{c=1}^C w_cp_c(x)
\end{equation}

Like a parametric distribution, a mixture distribution may be evaluated using
existing software like the \texttt{distr} package in \texttt{R} 
\cite{camphausen2007distr}. 
And 
scoring may be done  
using the LogS, CRPS, and IS. A mixture distribution, like its parametric 
distribution components, has an infinite resolution.
A mixture distribution may be more flexible than a single component 
parametric distribution in 
terms of distribution shape. According to 
McLachlan and Peel, a mixture of 
normal densities with common variance may be used to approximate arbitrarily 
well
any continuous distribution \cite{peel2000finite} (see also 
\cite{nguyen2019approximations}). Thus, for an unconventional probability
distribution --such as an MCMC posterior sample-- it may be reasonable to 
approximate the distribution by fitting those samples to a mixture of normal
distributions.
Depending on the number of components a forecaster includes in a mixture 
forecast, the amount of storage per forecast might be as little as for a 
parametric forecast or as much as is permitted in the specific 
collaborative forecast project. 

An ensemble model may be constructed by using \eqref{eq:bma} only replacing 
$p_m$
with $p_m^M$ from \eqref{eq:dmd}. Solving for weights may also be done by 
maximizing
the likelihood of the forecast or minimizing the CRPS. However with the 
added complexity of component models being mixture distributions the 
computation is likely to be more expensive. An example where this is
true is when minimizing the CRPS when the exact mixture distribution does not 
produce a closed form CRPS \cite{baran2018combining}. In large projects like
the COVID-19 Forecast Hub, if an equal weight is not assigned to each component,
it may be determined that models not reaching a certain standard of predictive
performance are
assigned an ensemble weight of 0. This would simplify an ensemble model to 
include only the best performing forecasts.

Table \ref{table:repscomp} shows how a mixture distribution forecast 
compares with the other formats discussed in terms of methods for scoring,
information and resolution provided, methods for ensemble building, and computer
storage requirement. To summarize, a continuous mixture distribution has the 
infinite resolution of a parametric distribution with the flexibility of a bin
distribution, a sample distribution, and a set of quantiles. The common proper
scoring rules LogS and CRPS may be used to score a mixture forecast. 
The storage requirement is
comparable to that of a bin distribution or a set of quantiles. And MA may be 
used for building an ensemble.
In Section
\ref{section:conmixforc} we show how a mixture distribution may be constructed,
scored, and used to construct an ensemble using software available in 
\texttt{R}.

\begin{table}

  \caption[Forecast representation comparison]{This table compares scoring,
     information, ensemble building, and storage requirements for the different
     forecast representations discussed. To summarize 
     a continuous mixture distribution has the 
  infinite resolution of a parametric distribution with the flexibility of a bin
  distribution, a sample distribution, and a set of quantiles. The common proper
  scoring rules LogS and CRPS may be used to score a mixture forecast. 
  The storage requirement is
  comparable to that of a bin distribution or a set of quantiles. And MA may be 
  used for building an ensemble.}

  \begin{tabular*}{\textwidth}{@{\extracolsep{\fill}} 
      l 
      *{8}{c} 
      S[table-format=4.0] }
    \toprule
    Type 
    & \multicolumn{3}{c}{Scoring} 
    & \multicolumn{3}{c}{Ensemble} 
    & Resolution
    & {Storage} 
    & {Shape/Flexibility} \\  
    \cmidrule{2-4}  \cmidrule{5-7}
    & LogS & CRPS & IS & MA & QA & Sample & & \\
    \midrule
    Bins & x & x & x & x & x & x & \# bins & 100s & {limited by binning scheme} \\
    Quantiles & & & x & & x & & \# quantiles & 10s & {unknown shape, no tail info} \\
    Parametric & x & x & x & x & x & x & $\infty$ & 10 & {well known distributions} \\
    Sample & & x & x & & x & x & $n$ draws & 1000 & {flexible shape} \\
    Mixture & x & x & x & x & x & x & $\infty$ & 10s & {flexible shape}\\
    \bottomrule
  \end{tabular*}
  \label{table:repscomp}

\end{table}


\section{Mixture distributions in a collaborative forecast project}
\label{section:conmixforc}

The CDC flu competition and the COVID-19 Forecast Hub as well as other 
collaborative projects
have their own established systems for receiving, 
scoring, and constructing ensemble forecasts. A transition from using bin or 
quantile forecasts to using mixture distribution forecasts
would require
a few adjustments to those systems. In this section we outline how some of these
adjustments may be implemented. We also present tools which may be used
to build forecasts from submissions, score those forecasts, and construct 
ensembles from them.

\subsection{Submission format}
For a collaborative forecast project to run smoothly, forecast submissions from 
all forecasters 
should follow the same format. For both the CDC flu competition and
the COVID-19 Forecast Hub, teams provide a \texttt{.csv} spreadsheet which 
contains the distributional information for one or multiple forecasts. Tables
\ref{table:dstan} and \ref{table:qstan} show what variables are included in 
those submissions and a
couple rows to illustrate possible values.
The column variables include \texttt{location}, \texttt{target}, \texttt{type},
\texttt{unit}, \texttt{bin} or \texttt{quantile}, and \texttt{value}.
Here \texttt{location} defines the specific county, state, or country of the 
forecast. The \texttt{target} variable defines what is forecast with levels:
season onset, deaths, hospitalizations, etc. The \texttt{type} variable defines
the type for the \texttt{value} variable with levels of point, 
bin, or quantile.
The \texttt{unit} variable defines the time frame of the forecast with levels of
one week,
two weeks, four weeks, etc. The variables \texttt{bin} and \texttt{quantile}
give a specific bin or a specific quantile. The \texttt{value} variable is a 
number that 
either gives the probability associated with a bin or the value associated 
with a quantile.

A single submission may include many forecasts aimed at forecasting 
different combinations of \texttt{location}, \texttt{target}, and \texttt{unit}. 
A set of rows which share
the same specific combination of \texttt{location}, \texttt{target}, and 
\texttt{unit} constitute a single forecast. One forecast for the CDC flu
competition may require up to 131 rows whereas in the COVID-19 Forecast Hub one 
forecast may require up to 23 rows.

\begin{table}[h!]
\begin{center}
\begin{minipage}{10cm}
\centering
\caption[Influenza competition submission example]{This table 
 shows a few rows of a 
 submission file
 for a bin forecast like those in the CDC flu 
 competition. A set of rows which share the same combination of 
 \texttt{location}, \texttt{target}, and \texttt{unit} make up a
 single forecast. One submission may include many forecasts specified by 
 differing combinations of those three columns.}
 \begin{tabular}{|c|c|c|c|c|c|}
 \hline
    location & target & type & unit & bin & value  \\ \hline
    us national & season onset & bin & week & 0.0 & . \\
    us national & season onset & bin & week & 0.1 & . \\
    ... & ... & ... & ... & ... & ... \\
 \hline
 \end{tabular}
 
 \label{table:dstan}
 \end{minipage}
 \end{center}
\end{table}

\begin{table}[h!]
\begin{center}
\begin{minipage}{11cm}
\centering
\caption[COVID-19 Forecast Hub competition submission example]{This shows
 a few rows of a 
 submission file
 for a quantile forecast like those in the COVID-19 Forecast Hub. 
 A set of rows which share the same combination of 
 \texttt{location}, \texttt{target}, and \texttt{unit} make up a
 single forecast. One submission may include many forecasts specified by 
 differing combinations of those three columns.}
 \begin{tabular}{|c|c|c|c|c|c|}
 \hline
    location & target & type & unit & quantile & value  \\ \hline
    us national & season onset & quantile & week & 0.01 & . \\
    us national & season onset & quantile & week & 0.025 & . \\
    ... & ... & ... & ... & ... & ... \\
 \hline
 \end{tabular}
 
 \label{table:qstan}
 \end{minipage}
 \end{center}
\end{table}

Table \ref{table:mstan} illustrates adjustments made to the submission formats
from 
Tables \ref{table:dstan} and \ref{table:qstan} which make a usable submission 
format for mixture distribution forecasts. In such a format, each row represents
one component distribution used in a mixture distribution.
The variables \texttt{bin} or \texttt{quantile} and \texttt{value} are removed 
and replaced with \texttt{family},
\texttt{param1}, \texttt{param2}, and \texttt{weight} where \texttt{family} is 
the distribution family of the component,
\texttt{param1} and \texttt{param2} are the parameters for the component 
distribution, and \texttt{weight} is the 
weight $w_i$ for the $i^{th}$ component.

\begin{table}[h!]
\centering
\caption[Mixture distribution forecast submission example]{This is an example 
 of a 
 submission file
 for a disease outbreak forecast using a mixture distribution representation.
 Each row represents a component distribution
 of a mixture distribution. The variables
 location, target, and unit specify what is being forecast. The variable type
 specifies that the row represents a parametric distribution. The variables
 family, param1, and param2 specify the exact component distribution. 
 And the variable 
 weight specifies the weight $w_i$ that the $i^{th}$ component distribution is
 assigned in the
 mixture distribution. Here two components are shown with distributions
 Normal$(a_n, b_n)$, Lognormal$(a_l, b_l)$ and weights $w_1$ and $w_2$.}
 \begin{tabular}{|c|c|c|c|c|c|c|c|c|}
  
 \hline
    location & target & type & unit & family & param1 & param2 & weight
    \\ \hline
    us national & season onset & dist & week & norm & $a_n$ & $b_n$ & $w_1$\\
    us national & season onset & dist & week & lnorm & $a_l$ & $b_l$ & $w_2$\\
    ... & ... & ... & ... & ... & ... & ... & ...\\
 \hline
 \end{tabular}
 \begin{minipage}{11cm}

 \label{table:mstan}
 \end{minipage}
\end{table}

For reasons of storage and computation, a forecast project may have a limit to 
the number of components allowed per forecast.
For reference, a mixture distribution forecast following the 
format in table \ref{table:dstan} with 17
components would require $17 \times 8 = 136$ pieces of information
submitted per forecast. 
A submission to the COVID-19 Forecast Hub forecast with 23 quantiles
according to the format in Table \ref{table:dstan} requires
$23 \times 6 = 138$ cells. Thus if the COVID-19 Forecast Hub were to change the
forecast representation from quantile forecasts to mixture forecasts but 
continue allowing the same amount of data per forecast, 
a mixture distribution with 17 components could be used in one forecast. That
many components could allow for a large range of distribution shapes and 
flexible forecasts.

In the remainder of this section, explanations of how to work with 
mixture distributions submitted according to Table \ref{table:mstan} are 
given. Also given is
\texttt{R} code which 
demonstrates constructing a mixture distribution from a forecast submission, 
scoring the forecast, and building an ensemble from two separate submissions.

\subsection{Mixture construction and scoring tools}
\label{section:tools}

A single \texttt{.csv} submission file of the format in Table \ref{table:mstan} 
may contain multiple forecasts forecasting different combinations of location,
target, and unit. Selecting only rows which share a specific combination of 
location, target, and unit will produce a table representing a single forecast.
That table may look like Tables \ref{tab:preddf1} and 
\ref{tab:preddf2}. If the table is saved as a standard \texttt{data.frame} in 
\texttt{R}, then tools based on the \texttt{distr} package 
\cite{camphausen2007distr} may be used for evaluating a mixture distribution 
with the component distributions in the table.

The \texttt{distr} package contains a function 
\texttt{UnivarMixingDistribution()} which takes as arguments a list of 
distributions
and a vector of weights for each distribution and an object of class 
\texttt{AbscontDistribution} is returned. An \texttt{AbscontDistribution} class
is a mother class which defines a random number generator, pdf,
CDF, and quantile function for continuous distributions from common
families contained in the \texttt{distr} package
and for mixture distributions with 
component distributions from those families.
We wrote a function \texttt{MakeDist()} (see APPENDIX)
which takes on a \texttt{data.frame} with variables
family, param1, param2, param3, and weight and where each row represents a
component distribution in a mixture distribution. The function 
\texttt{MakeDist()} calls on the \texttt{UnivarMixingDistribution()} function
and returns a mixture distribution object of class \texttt{AbscontDistribution}.

If a forecast such as in Table \ref{tab:preddf1} is taken as an argument in 
\texttt{MakeDist()}, the resulting mixture distribution may then be evaluated 
with functions for the pdf, CDF, quantile function, and random samples from the
mixture distribution may be drawn. The distribution may then be scored using the
LogS or CRPS.

\begin{table}[h!]
\centering
\caption[Illustrative forecast 1]{This is an illustrative example of a
 mixture distribution
 forecast where the distribution is described in a data frame. The first 
 component is a Lognormal$(2,1)$ with a weight of 0.3 in the mixture and the 
 second component is a Normal$(2.1,1)$ with a weight of 0.7. The distribution
 family abbreviations are capitalized here because that is how they will be
 requested in the \texttt{MakeDist()} function. Refer to Table
 \ref{table:mixfams} in the APPENDIX for more details. }
 \begin{tabular}{|c|c|c|c|c|}
 \hline
    family & param1 & param2 & param3 & weight
    \\ \hline
    Lnorm & 2 & 1 & NA & 0.3  \\
    Norm & 2.1 & 1 & NA & 0.7 \\
 \hline
 \end{tabular}
  \begin{minipage}{11cm}
 
 \label{tab:preddf1}
 \end{minipage}
\end{table}

\begin{table}[h!]
\centering
\caption[Illustrative forecast 2]{This is a second
 illustrative example of a mixture distribution
 forecast where the distribution is described in a data frame. The first 
 component is a Normal$(1.5,1)$ with a weight of 0.4 in the mixture and the 
 second component is a Normal$(4,2)$ with a weight of 0.6.
 The distribution
 family abbreviations are capitalized here because that is how they will be
 requested in the \texttt{MakeDist()} function. Refer to Table
 \ref{table:mixfams} in the APPENDIX for more details.}
 \begin{tabular}{|c|c|c|c|c|}
 \hline
    family & param1 & param2 & param3 & weight
    \\ \hline
    Norm & 1.5 & 1 & NA & 0.4  \\
    Norm & 4 & 2 & NA & 0.6 \\
 \hline
 \end{tabular}
  \begin{minipage}{11cm}
 
 \label{tab:preddf2}
 \end{minipage}
\end{table}

\begin{figure}[htbp]
\centerline{\includegraphics[scale=.9]{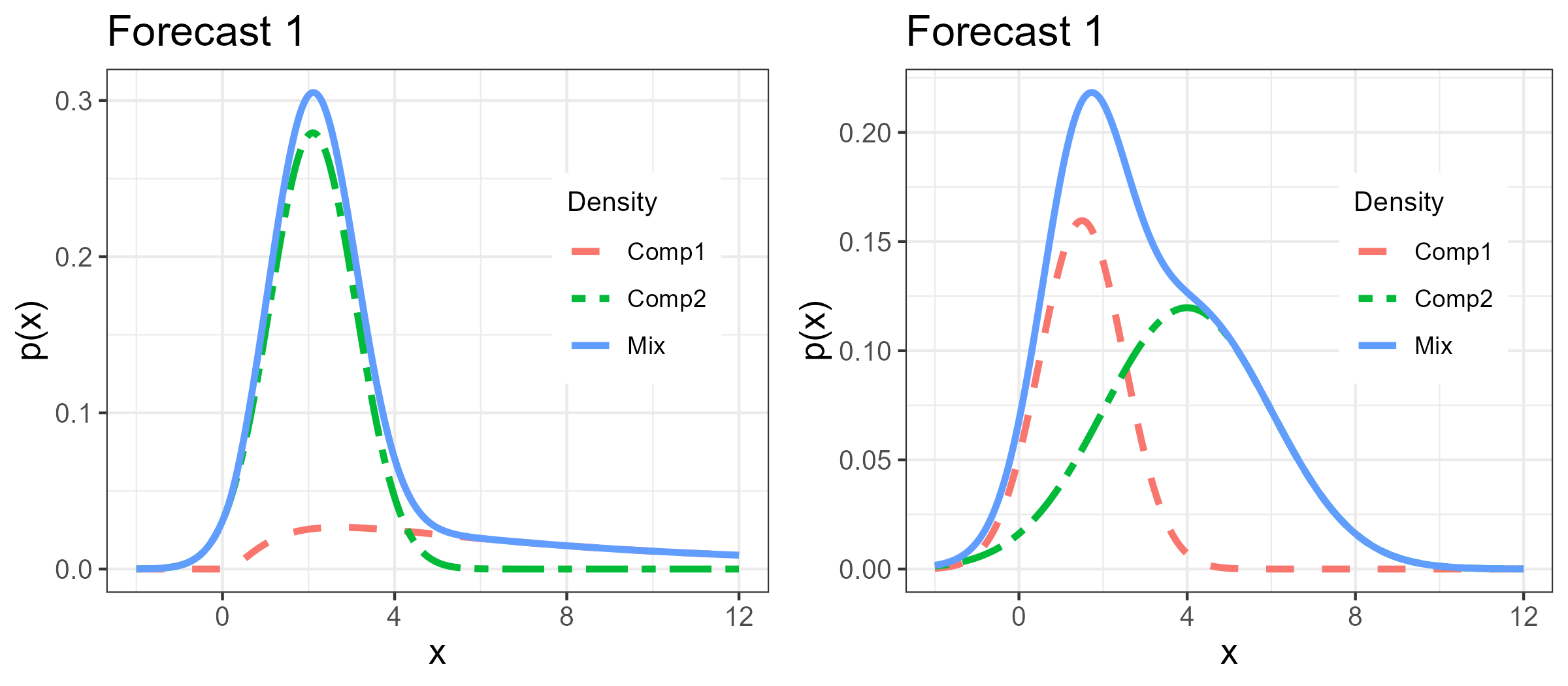}}
\begin{center}
 \begin{minipage}{11cm}
\caption[Example mixture distributions]{These plots show the density functions 
of
the two example mixture distribution forecasts along with the component density
functions scaled by the corresponding weights}
\label{fig:toymixs}
\end{minipage}
\end{center}
\end{figure}

Here we include code to illustrate the process of constructing and scoring two 
separate forecasts. We suppose that Table \ref{tab:preddf1} represents a 
submitted forecast from one forecaster and Table \ref{tab:preddf2} represents
a forecast of the same event from a second forecaster. Table \ref{fig:toymixs}
shows plots of the pdfs for both mixture forecasts.
Note the additional \texttt{param3} variable
in Tables \ref{tab:preddf1} and \ref{tab:preddf1}. This variable is included
in the table because of the functionality of the \texttt{MakeDist()} function
which allows for component distributions of up to three parameters.
The code here shows these two forecasts as \texttt{data.frame}s and how the 
\texttt{MakeDist()} 
function is used to create the distributions in \texttt{R}. Once the 
distributions are created as \texttt{AbscontDistribution} objects, then 
functions for evaluating a pdf and a CDF for each are created.

\begin{knitrout}
\definecolor{shadecolor}{rgb}{0.969, 0.969, 0.969}\color{fgcolor}\begin{kframe}
\begin{alltt}
\hlstd{preddf1}
\end{alltt}
\begin{verbatim}
##   family param1 param2 param3 weights
## 1  Lnorm    2.0      1     NA     0.3
## 2   Norm    2.1      1     NA     0.7
\end{verbatim}
\begin{alltt}
\hlstd{preddf2}
\end{alltt}
\begin{verbatim}
##   family param1 param2 param3 weights
## 1   Norm    1.5      1     NA     0.4
## 2   Norm    4.0      2     NA     0.6
\end{verbatim}
\begin{alltt}
\hlcom{#make mixture distributions from prediction submissions}
\hlstd{mdist1} \hlkwb{<-} \hlkwd{MakeDist}\hlstd{(preddf1)}
\hlstd{mdist2} \hlkwb{<-} \hlkwd{MakeDist}\hlstd{(preddf2)}
\end{alltt}
\end{kframe}
\end{knitrout}

\begin{knitrout}
\definecolor{shadecolor}{rgb}{0.969, 0.969, 0.969}\color{fgcolor}\begin{kframe}
\begin{alltt}
\hlcom{#make pdfs for mixture predictions}
\hlstd{dmdist1} \hlkwb{<-} \hlkwa{function}\hlstd{(}\hlkwc{x}\hlstd{) \{distr}\hlopt{::}\hlkwd{d}\hlstd{(mdist1)(x)\}}
\hlstd{dmdist2} \hlkwb{<-} \hlkwa{function}\hlstd{(}\hlkwc{x}\hlstd{) \{distr}\hlopt{::}\hlkwd{d}\hlstd{(mdist2)(x)\}}

\hlcom{#make cdfs for mixture predictions}
\hlstd{pmdist1} \hlkwb{<-} \hlkwa{function}\hlstd{(}\hlkwc{x}\hlstd{) \{distr}\hlopt{::}\hlkwd{p}\hlstd{(mdist1)(x)\}}
\hlstd{pmdist2} \hlkwb{<-} \hlkwa{function}\hlstd{(}\hlkwc{x}\hlstd{) \{distr}\hlopt{::}\hlkwd{p}\hlstd{(mdist2)(x)\}}
\end{alltt}
\end{kframe}
\end{knitrout}

The LogS or the CRPS may then be calculated for each forecast using the pdf
and CDF functions respectively. Here will will assume that the true value
which both forecasts attempted to predict was \texttt{3}. 
The \texttt{CRPS()} function
here is one that we wrote and is
included in the APPENDIX. It is seen in the code below that under the
LogS, forecast 1 from Table \ref{tab:preddf1} outperforms forecast 2 from Table
\ref{tab:preddf2} with scores of 1.547 and 
1.849 respectively. 
However, under the CRPS, forecast 2 outperforms forecast 1
with scores 0.635 and 
0.635 respectively. 
We continue to use these same
forecasts in Section \ref{section:enscon} used in constructing
an ensemble forecast.

\begin{knitrout}
\definecolor{shadecolor}{rgb}{0.969, 0.969, 0.969}\color{fgcolor}\begin{kframe}
\begin{alltt}
\hlcom{#realized observation}
\hlstd{xstar} \hlkwb{<-} \hlnum{3}

\hlcom{#LogS for predictions at the realized observation}
\hlopt{-}\hlkwd{log}\hlstd{(}\hlkwd{dmdist1}\hlstd{(xstar))}
\end{alltt}
\begin{verbatim}
## [1] 1.547238
\end{verbatim}
\begin{alltt}
\hlopt{-}\hlkwd{log}\hlstd{(}\hlkwd{dmdist2}\hlstd{(xstar))}
\end{alltt}
\begin{verbatim}
## [1] 1.848796
\end{verbatim}
\end{kframe}
\end{knitrout}

\begin{knitrout}
\definecolor{shadecolor}{rgb}{0.969, 0.969, 0.969}\color{fgcolor}\begin{kframe}
\begin{alltt}
\hlcom{#CRPS for predictions at the realized observation}
\hlkwd{CRPS}\hlstd{(pmdist1,}\hlkwc{y}\hlstd{=xstar)}
\end{alltt}
\begin{verbatim}
## [1] 0.6348212
\end{verbatim}
\begin{alltt}
\hlkwd{CRPS}\hlstd{(pmdist2,}\hlkwc{y}\hlstd{=xstar)}
\end{alltt}
\begin{verbatim}
## [1] 0.5306083
\end{verbatim}
\end{kframe}
\end{knitrout}

\subsection{Ensemble construction}
\label{section:enscon}

To construct an ensemble distribution from multiple mixture distributions, the 
\texttt{UnivarMixingDistribution()} function may be used. The function takes 
two or more
\texttt{AbscontDistribution} distribution objects, including mixture 
distribution objects, and a vector of weights 
corresponding to each object. A new \texttt{AbscontDistribution} object is 
returned as an ensemble of mixture distributions as in (\ref{eq:bma}). 
Since they are \texttt{AbscontDistribution} objects, \texttt{mdist1} and 
\texttt{mdist2} created in the code
in Section \ref{section:tools} may be input as arguments into the function
\texttt{UnivarMixingDistribution()}, but weights for each object also need to
be determined.

At the onset of a collaborative forecast before there are true event 
observations which the forecasts may be scored on, 
it may make sense to assign an equal weight to each 
component distribution in an ensemble. As a project progresses, however, 
assigning weights based on past performance may be desired. As mentioned in 
section \ref{section:pardist}, weights may be selected by maximizing the
likelihood of (\ref{eq:bma}) or by minimizing the CRPS. Another method of 
selecting weights is to use the posterior model probability. 

If we have $T$ models $M_1, ..., M_T$ the posterior model probability of $M_t$ is 
defined as in \eqref{eq:pmp} where $p(\cdot |M_t) := p_t(\cdot)$ is the pdf 
of the model distribution and 
$p(M_t)$ is the prior probability assigned to the model. A common approach is to
assume the prior
probabilities for each model are equal or $p(M_t) = 1/T$ for all $t$ in which 
case
(\ref{eq:pmp}) is reduced to \eqref{eq:pmpeq}. In this case the posterior model 
probability for the $t^{th}$ model is equal to the exponential of its negative 
LogS or
$p(M_t|x) = e^{-\text{LogS}(p(M_t|x))}$, so the performance of a forecast 
based on the
LogS is directly related to its posterior model probability and may be used 
as an 
ensemble weight. For an observed event $x^*$, ensemble weights $(w_t)$ from 
\eqref{eq:bma} may be defined as $w_t := p(M_t|x^*)$.

\begin{equation}
\label{eq:pmp}
p(M_t|x) = \frac{p(x|M_t)p(M_t)}{p(x)}
         = \frac{p(x|M_t)p(M_t)}{\sum_{k=1}^T p(x|M_k)p(M_k)}
\end{equation}

\begin{equation}
\label{eq:pmpeq}
p(M_t|x) =  \frac{p(x|M_t)}{\sum_{k=1}^Tp(x|M_k)}
\end{equation}

\begin{figure}[htbp]
\centerline{\includegraphics[scale=1]{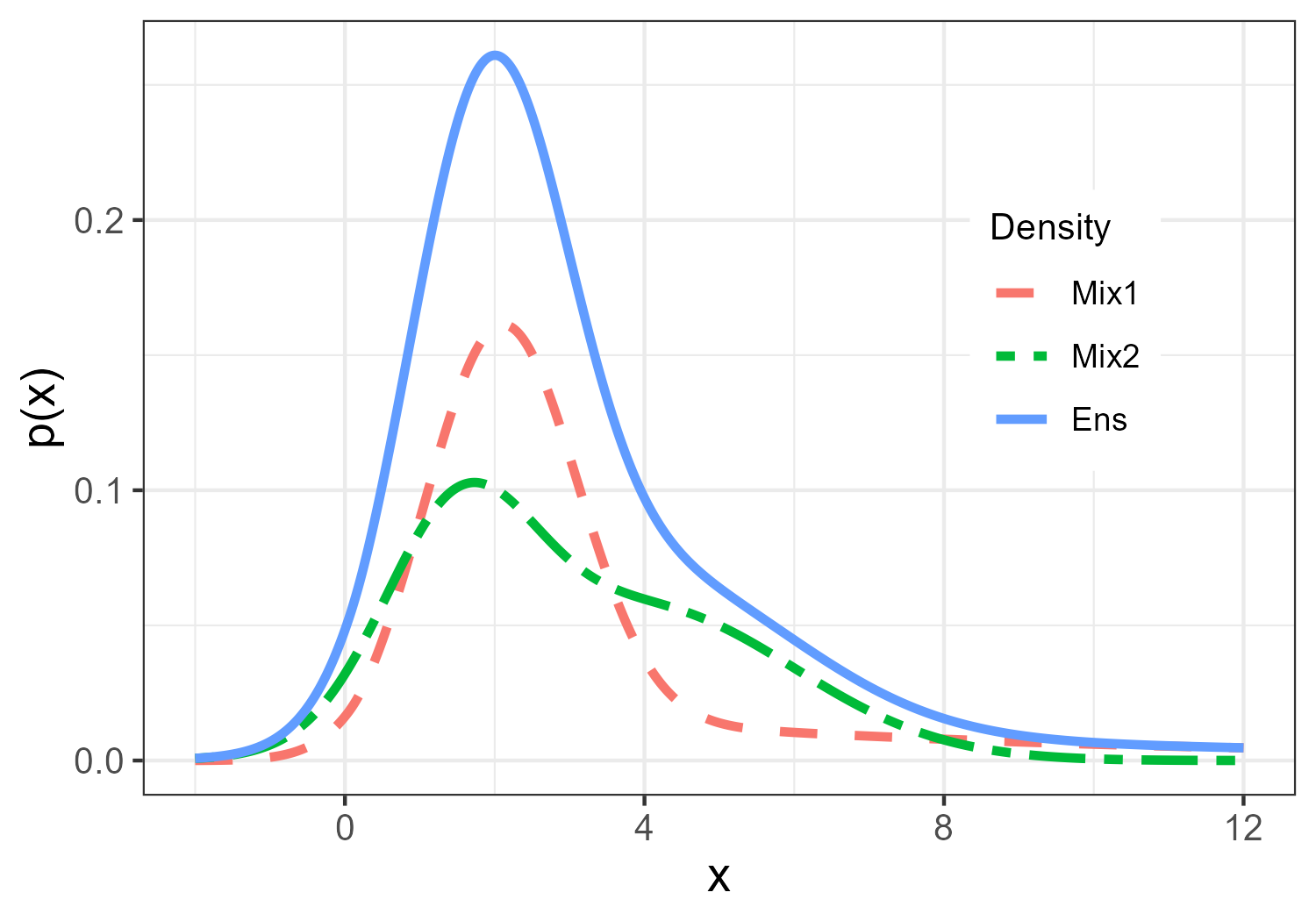}}
\begin{center}
 \begin{minipage}{10cm}
\caption[Example ensemble forecast]{Ensemble forecast made from forecast 1 and 
forecast 2 from Section \ref{section:tools}. The red line is the density 
component of mixture forecast 1 with weight 0.529. The 
green line is the density component of mixture forecast 2 with weight 
0.471. The blue line is the density of the ensemble forecast.}
\label{fig:mixense}
\end{minipage}
\end{center}
\end{figure}

Using the illustrative example from Section \ref{section:tools}, 
the following code shows how
to use the posterior model probability to select weights, construct an 
ensemble distribution, and score the ensemble forecast. Here again we take the 
true event value to be \texttt{3}.
The ensemble distribution
along with component distributions is shown in Figure \ref{fig:mixense}.

\begin{knitrout}
\definecolor{shadecolor}{rgb}{0.969, 0.969, 0.969}\color{fgcolor}\begin{kframe}
\begin{alltt}
\hlcom{#posterior model probability for calculating weights}
\hlstd{w1} \hlkwb{<-} \hlkwd{pmdist1}\hlstd{(xstar)}\hlopt{/}\hlstd{(}\hlkwd{pmdist1}\hlstd{(xstar)} \hlopt{+} \hlkwd{pmdist2}\hlstd{(xstar))}
\hlstd{w2} \hlkwb{<-} \hlnum{1}\hlopt{-}\hlstd{w1}
\hlstd{w1}
\end{alltt}
\begin{verbatim}
## [1] 0.5286434
\end{verbatim}
\begin{alltt}
\hlstd{w2}
\end{alltt}
\begin{verbatim}
## [1] 0.4713566
\end{verbatim}
\begin{alltt}
\hlcom{#build ensemble with calculated weights}
\hlstd{ensdist} \hlkwb{<-} \hlstd{distr}\hlopt{::}\hlkwd{UnivarMixingDistribution}\hlstd{(mdist1,}
                                           \hlstd{mdist2,}
                                           \hlkwc{mixCoeff} \hlstd{=} \hlkwd{c}\hlstd{(w1,w2))}

\hlcom{#pdf and cdf for ensemble}
\hlstd{densdist} \hlkwb{<-} \hlkwa{function}\hlstd{(}\hlkwc{x}\hlstd{) \{(distr}\hlopt{::}\hlkwd{d}\hlstd{(ensdist)(x))\}}
\hlstd{pensdist} \hlkwb{<-} \hlkwa{function}\hlstd{(}\hlkwc{x}\hlstd{) \{(distr}\hlopt{::}\hlkwd{p}\hlstd{(ensdist)(x))\}}

\hlcom{#LogS for predictions at the realized observation}
\hlopt{-}\hlkwd{log}\hlstd{(}\hlkwd{densdist}\hlstd{(xstar))}
\end{alltt}
\begin{verbatim}
## [1] 1.678156
\end{verbatim}
\begin{alltt}
\hlcom{#CRPS for predictions at the realized observation}
\hlkwd{CRPS}\hlstd{(pensdist,}\hlkwc{y}\hlstd{=xstar)}
\end{alltt}
\begin{verbatim}
## [1] 0.5486368
\end{verbatim}
\end{kframe}
\end{knitrout}

\section{Retrospective analysis}
\label{section:retrostud}

For large collaborative forecast projects having already established the 
representation formats for forecasting, it may be difficult for
teams to adjust to using mixture distributions. There may be several
reasons for this, including that not all 
forecast modeling methods will produce forecasts which may conveniently be
represented by a mixture distribution.
In this section we attempt to assess whether or not bin forecasts from the CDC
flu competition or quantile forecasts from the COVID-19 Forecast Hub may be 
reasonably approximated by a mixture distribution with normal components as the 
number of components in the distribution increases.

Forecasters in both the CDC flu competition and the COVID-19 Forecast Hub do 
not include with their forecast submissions information about modeling methods 
or 
distributional assumptions. Thus the only information we have for fitting
distributions are bin
forecasts and quantile forecasts. We are unaware of formal statistical 
methods for fitting parametric distributions or mixture distributions to bin
distributions. Methods of fitting a distribution to quantiles 
include Bayesian Quantile Matching
\cite{nirwan2020bayesian}, step interpolation with exponential tails 
\cite{quinonero2005evaluating},
and the Method of Simulated Quantiles 
\cite{dominicy2013method}. These studies, however, lack claims that the 
methods
for fitting are statistically formal.
Nirwan and Bertshinger state that minimizing the mean square error between 
quantile values and a CDF function has been the most common way to fit a 
distribution
to a set of quantiles. This is the method we will use in Section 
\ref{section:covdretro}.
Because of the lack of statistically formal methods for fitting a parametric
distribution to a bin distribution or a set of quantiles, 
it should be noted that any conclusions made in 
this section may not be 
stated in terms of statistical certainty.

\subsection{CDC flu competition}
The CDC Retrospective Forecasts project on zoltrdata.com \cite{zoltarflu} 
contains
869,638 probabilistic influenza-like illness forecasts for all combinations
of 11 regions 
in the United States and seven targets from 27 different modeling teams. These
include forecasts made during all flu seasons between October 2010 and December
2018. All forecasts are represented by bin distributions.

To assess whether the bin probabilities may be more closely approximated by 
mixture distributions with an increasing number of components, 5 mixture
distributions with one to five normal components were fit to each of a selected set 
of bin forecasts.  
In fitting a distribution to a forecast, we want minimize \eqref{eq:kld}.
Equation
\eqref{eq:kld} is a variation of the Kullback Leibler divergence (KLD).
Here $\bm{p}$ represents a bin distribution where $p_i > 0$ is the reported
probability
for the $i^{th}$ bin $B_i := [b_{i-1}, b_i)$, and $K$ is the number of bins.
$M_{\theta}^C$ is a random variable of a mixture distribution with $C$ 
components and parameter
vector $\theta$.
The fitted parameter vector $\hat{\theta}$ is the solution to
\eqref{eq:discfit}.

\begin{equation}
  D(\bm{p}|M_{\theta}) = \sum_{i=1}^K p_i \text{log}\left(\frac{p_i}{P(M^C_{\theta}
  \in B_i)}\right)
  \label{eq:kld}
\end{equation}

\begin{equation}
\hat{\theta} = \underset{\theta}{\operatorname{argmin}}\sum_{i=1}^K p_i \text{log}\left(\frac{p_i}{P(M^C_{\theta}
  \in B_i)}\right)
\label{eq:discfit}
\end{equation}

Pulling forecasts from zoltardata.com and fitting mixture distributions to them
was computationally expensive, so we limited the analysis to 1,141 forecasts. To
select these forecasts, we sampled from the 869,638 forecasts as follows. 
For each submission -a single submission may contain multiple forecasts
for various units and targets forecasted- there is a recorded date corresponding
to the week of the forecast. There are 246 total dates.
The number of submissions for each week was counted
and 10 weeks were randomly selected with probabilities based on the number of
submissions by week. All submissions from the selected 10 weeks were pulled from
zoltardata.com, but the only forecasts kept for fitting were US national level
forecasts for 
1, 2, 3, and 4 week ahead and season peak percentage and forecasts with more 
4 bins. Twenty-six of twenty-seven teams were represented by the selected
forecasts.

With forecasts selected, we then fit five mixture distributions to each. For 
each distribution to be fit there are $C \in \{ 1,2,3,4,5 \}$ normal components.
In the mixture distribution there are $C$ mean values, a common standard 
deviation 
shared by each component, and $C$ component weights $\omega_1,...,\omega_C$. 
Thus
the parameter vector $\theta = (\mu_1,...,\mu_C,\sigma,\omega_1,...,\omega_C)$.

To ensure in the
optimization that $\omega_1,...,\omega_{C} > 0$ and
$\sum_{i=1}^{C} \omega_i = 1$ we optimize the parameters $\nu_1=0,...,\nu_{C}$
and set $\omega_i = \frac{e^{\nu_i}}{\sum_{i=1}^{c} e^{\nu_i}}$. Since
$\nu_1 = 0$, only the $C-1$ parameters $\nu_2,...,\nu_C$ parameters are
optimized.
To maintain order in the set of mixture components, we add the constraint
$\mu_1 < \mu_2 < ... < \mu_C$. This is enforced by taking
$\mu_1=\mu_1,\mu_2=\mu_1+e^{\alpha_1},...,\mu_C=\mu_{C-1}+e^{\alpha_{C-1}}$
so that the optimized parameters for means are 
$\mu_1,\alpha_1,...,\alpha_{C-1}$.
The necessary condition that $\sigma>0$ is enforced by setting
$\sigma = e^{\eta}$ and optimizing over $\eta$.
Thus the $2C$ parameter vector to be optimized is 
$\gamma = (\mu_1,\alpha_1,...,\alpha_{C-1},\eta,\nu_2,...\nu_C)$. 
The optimization was done iteratively by repeating the following steps.

1. Initialize
$\gamma_0 = (\mu_1^{(0)},\alpha_1^{(0)}...,\alpha_{C-1}^{(0)},\eta^{(0)},\nu_2^{(0)},...,\nu_C^{(0)})$

2. At step $m \geq 1$, set the following $2C$ parameter vectors where 

\begin{equation}
\notag \gamma_1^{(m)} = (\tilde{\mu_1}^{(m+1)},\alpha_1^{(m)}...,
\alpha_{C-1}^{(m)},\eta^{(m)},\nu_1^{(m)},...,\nu_C^{(m)})
\end{equation}

\begin{equation}
\notag \gamma_2^{(m)} = (\mu_1^{(m)},\tilde{\alpha_1}^{(m+1)}...,
\alpha_{C-1}^{(m)},\eta^{(m)},\nu_1^{(m)},...,\nu_C^{(m)})
\end{equation}

\begin{equation}
\notag \vdots
\end{equation}

\begin{equation}
\notag \gamma_{2C}^{(m)} = (\mu_1^{(m)},\alpha_1^{(m)}...,
\alpha_{C-1}^{(m)},\eta^{(m)},\nu_1^{(m)},...,\tilde{\nu_C}^{(m+1)})
\end{equation}

  where $\gamma_i^{(m)}$ is the parameter vector minimizing \eqref{eq:discfit}
  over the $i^{th}$ element while holding all other elements constant.

3. Set
$\gamma^{(m+1)} = \underset{\gamma_i^{(m)}}{\operatorname{argmin}} \sum_{i=1}^K p_i \text{log}\left(\frac{p_i}{P(M_{\gamma_i^{(m)}}
  \in B_i)}\right), i \in (1,...,2C)$
  
4. Return to step 2.

This process was run until 
$|D(p||M_{\theta^{(m+1)}} - D(p||M_{\gamma^{(m)}})|/ D(p||M_{\gamma^{(m)}}) < 0.001$
where $D(p||M_{\theta})$ is as defined in \eqref{eq:kld}, or $m = 500$. The 
optimization was done using the \texttt{optim} function in \texttt{R}, and the 
optimization algorithm used was either "BFGS" or "L-BFGS-B". Of the 1,141
selected forecasts, 1,103 were fit by the 5 mixture distributions. The remaining
forecasts were ignored in further analysis 
because the fitting algorithms failed to converge.

\begin{figure}[htbp]
\centerline{\includegraphics[scale=.84]{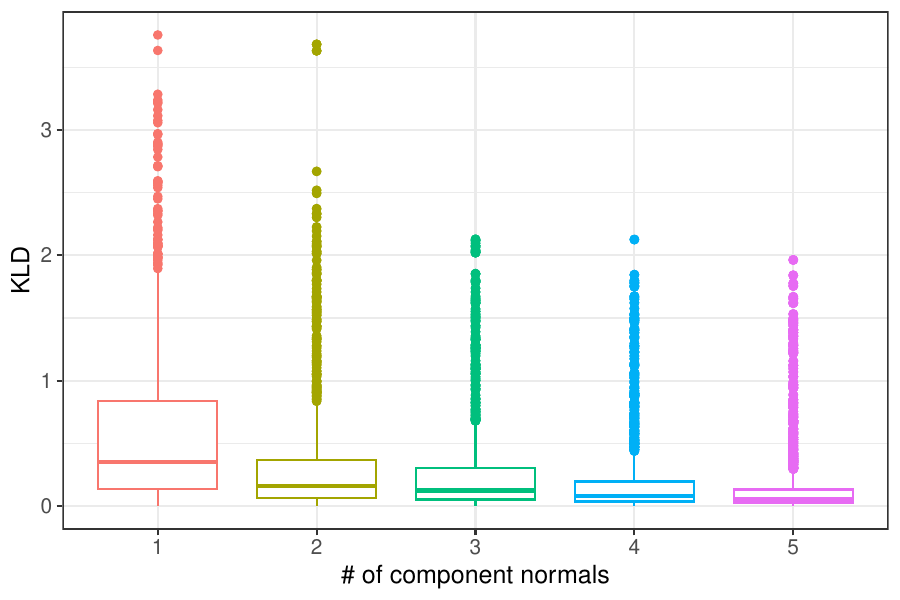}}
\begin{center}
\begin{minipage}{10cm}
\caption[Boxplots of KLD values for mixture distributions of increasing 
components fit to flu bin forecasts]{Each boxplot is made of the KLD from 
equation 
\eqref{eq:kld} 
calculated
for each sampled flu forecast. As the number of component normals fit to the
forecasts increases, the median values and spread decrease.}
\label{fig:klbox}
\end{minipage}
\end{center}
\end{figure}

\begin{figure}[htbp]
\centerline{\includegraphics[scale=.8]{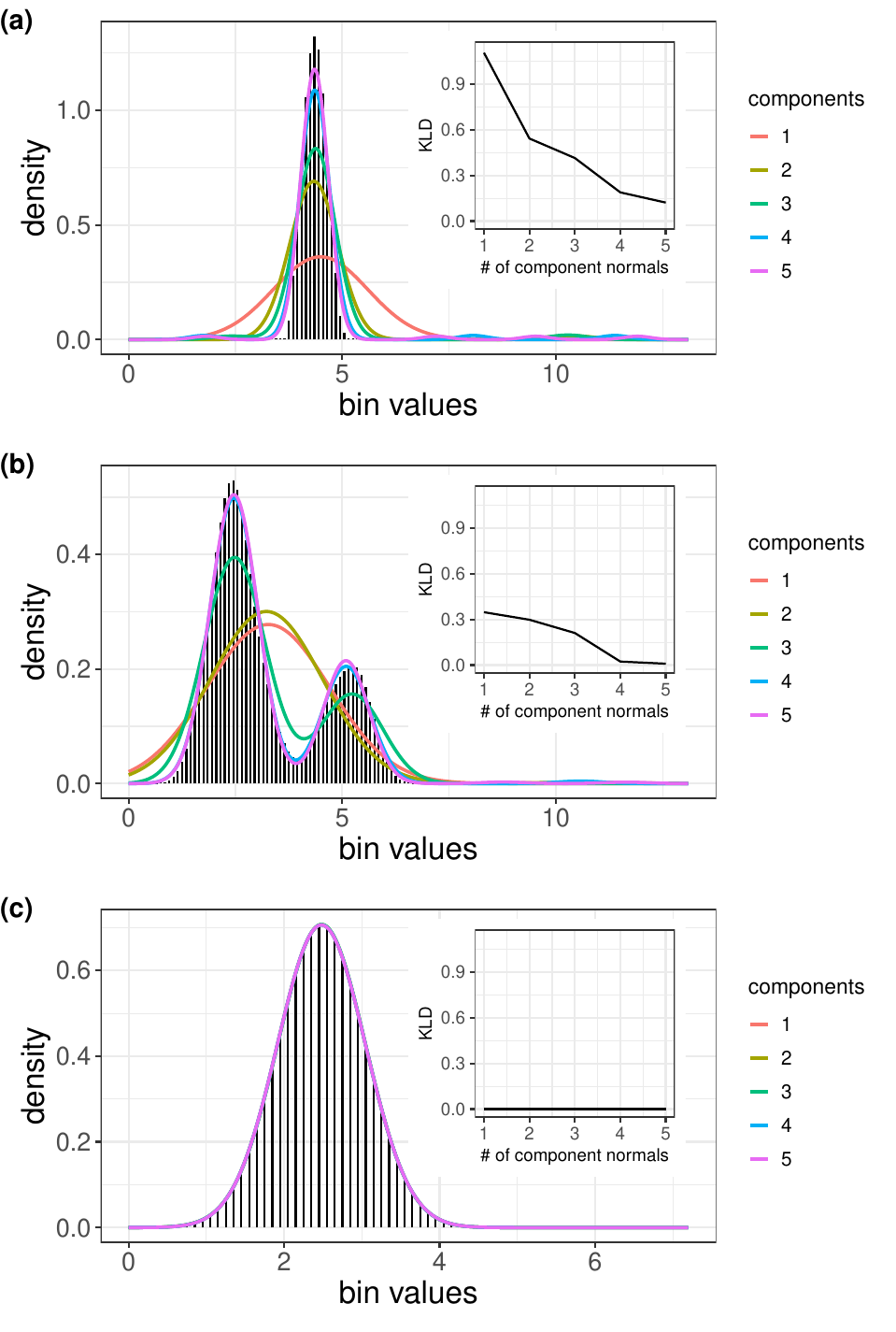}}
\begin{center}
\begin{minipage}{10cm}
\caption[Flu bin forecasts with fitted mixture pdfs and KLD values]{Three 
selected flu forecasts with the pdfs of mixture
distributions of one to five components plotted on top. The inner plot shows
the KLD value as the number of components increases. The bin probabilities
are multiplied by 10 to ensure they are on the same scale as the pdfs in the
plot.}
\label{fig:denf1}
\end{minipage}
\end{center}
\end{figure}

Figure 
\ref{fig:klbox} shows boxplots of KLD for all fit mixture distributions for 1 to
5 components. In general, the KLD between the actual forecast distribution and
the fit distribution tends to decrease as the number components in the fit 
mixture distribution increases. Figure \ref{fig:denf1} shows examples of density
functions of one to five components fit to a forecast.
The outer plots show fit mixture density functions plotted over the bin
probabilities where the probabilities are multiplied by 10 to give the 
same scale as the densities. The inner plots show the KLD of the forecast
and the fit distribution by number of components in the mixture distribution.

\begin{figure}[htbp]
\centerline{\includegraphics[scale=.8]{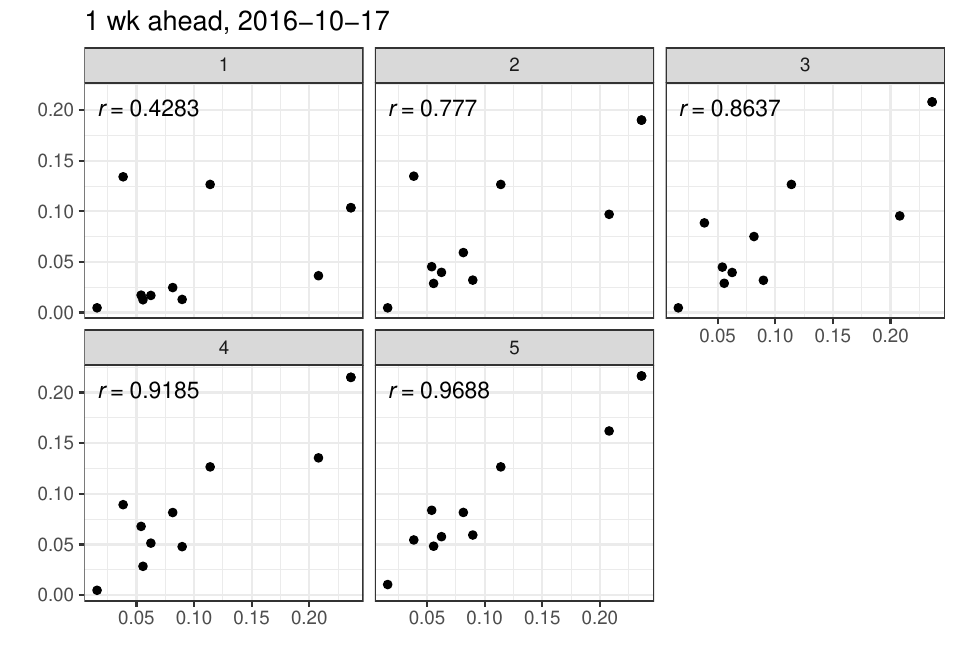}}
\begin{center}
\begin{minipage}{10cm}
\caption[Scatterplot of true probability scores vs. fit mixture probability 
scores]{Scatterplots with $p_t$ on the x-axis and $p_t^{M^C}$ on the y-axis
for all selected 1 week ahead forecasts for the date 10/17/16. The correlation
between the probabilities increases as the number of components in the fit 
mixture increases suggesting that the forecasts may be more closely approximated
as the number of components increases.}
\label{fig:prscat}
\end{minipage}
\end{center}
\end{figure}

To further compare the fits to the forecasts, we compared the
actual forecast performance of the bin forecast to the fit mixture 
distributions. For each week and target represented in the sample of 
selected forecasts, truth data was obtained. For each bin forecast, the bin 
in which the true value fell, 
$B_t$ was determined and the probability value in that bin
was noted as $p_t$. Then for each mixture distribution fit to that density, the 
probability $p^{M^C}_t = P(M^C \in B_t)$ within the true bin was calculated. 
All forecasts
were then classed by the specific target and week for which they were
forecasting. There were 50 total combinations of target and week. 
Figure \ref{fig:prscat} shows 5 scatterplots where the probabilities $p_t$ for
for 1 week ahead forecasts on 2016-10-17 are plotted against the probabilites
$p^{M^C}_t$. Each plot represents a different number of component distrubtions
in the mixture distribution fit. The linear correlation coefficient between the
probabilites is also given, and the correlation increases
as the number of components increases. Table \ref{table:probcor} shows the same
correlation trend for 5 different target/week combinations and for all forecasts
not broken out by target/week combinations.

\begin{table}[h!]
\begin{center}
\begin{minipage}{10cm}
\centering
\caption[Correlations of actual probability scores vs. fit mixture probability
  scores]{(a) 10/28/13 Season peak, (b) 12/2/13 4 wk, (c) 10/17/16 season
             peak, (d) 4/20/15 4 wk, (e) 4/20/15 1 wk. The values are the 
             correlations between the actual probability score from the 
             submitted forecasts and the probability according to the fit
             mixture distributions.}
 \begin{tabular}{c|c c c c c c} 
 Components   & Overall  & (a)     & (b)     & (c)     & (d)    & (e)    \\
 \hline
   1          & 0.5811   & 0.7608  & 0.789   & 0.4454  & 0.748  & 0.4752 \\
   2          & 0.738    & 0.9428  & 0.8885  & 0.6349  & 0.8025 & 0.8264 \\
   3          & 0.8341   & 0.9567  & 0.9626  & 0.7610  & 0.8271 & 0.8503 \\
   4          & 0.8445   & 0.978   & 0.9715  & 0.9428  & 0.7874 & 0.9114 \\
   5          & 0.8920   & 0.9844  & 0.9903  & 0.9691  & 0.7188 & 0.9454 \\
 \hline

 \end{tabular}
  
\label{table:probcor}
\end{minipage}
\end{center}
\end{table}

\subsection{COVID-19 Forecast Hub}
\label{section:covdretro}

As of October 2022 there were over 100 million forecasts  
from 122 different modeling teams submitted to the COVID-19 Forecast Hub 
\cite{zoltarcovid}. 
These 
forecasts covered all combinations of 3,202 municipalities (mostly counties) 
in the United States with 441 target/unit combinations. 
The first of these forecasts was 
submitted in March of 2020 shortly after the initial outbreak of the COVID-19 
virus in 
the US, and forecasts have been received weekly since then. 
The forecasts are all quantile forecasts made up of three or 11 predictive 
intervals --depending on the specific unit for the forecast--
and a median. Thus each forecast includes seven or 23 quantiles.

As in the retrospective study of influenza forecasts, we wanted to assess 
whether or not we can more closely approximate the quantile forecasts by 
increasing the number of components in a mixture distribution. Fitting a mixture
distribution to a set of quantiles was done in the same manner as to bin 
probabilities, only by choosing the parameter $\hat{\theta}$ which minimizes
the sum of square differences (SS) 
between the quantiles and the CDF of a mixture distribution as in equation
\eqref{eq:sse}. Here $\hat{\theta}$ is the solution to equation 
\eqref{eq:minthet}. In equation \eqref{eq:sse}, 
$q_i$ is the $i^{th}$ quantile (out of $N$ quantiles) from a forecast and
$F(\alpha_i|\hat{\theta})$ is a fit CDF evaluated at the 
value $\alpha_i$ with parameter $\hat{\theta}$.  
The parameter $\hat{\theta}$ is the solution to \eqref{eq:minthet}. 

\begin{equation}
  \text{SS} = \sum_{i=1}^m (q_i - F(\alpha_i| \hat{\theta}))^2
  \label{eq:sse}
\end{equation}

\begin{equation}
  \hat{\theta} = \arg\min_{\theta}
  \sum_{i=1}^m (\alpha_i - F(q_i| \theta))^2
  \label{eq:minthet}
\end{equation}

We randomly selected 9 weeks from which to pull forecasts from zoltardata.com
by searching through all teams which submitted any forecast. Forecasts
submitted between a Monday and the following Tuesday were considered to have 
come from the same week \cite{cdcflusightensemble}. 
The 9 weeks were selected randomly with probability 
relative to the total number of forecasts submitted that week. From those weeks, 
only US national forecasts for increasing and cumulative deaths from one to four 
weeks ahead
were selected for fitting. In total, 2,676 forecasts were selected. 
However, as in the influenza analysis, computational issues
made it difficult to fit five different mixture distributions to all quantile
forecasts, as well truth data was missing for two targets during one week in
September 2022. 
Thus the remainder of the anlysis included 2,319 forecasts. 

\begin{figure}[htbp]
\centerline{\includegraphics[scale=.8]{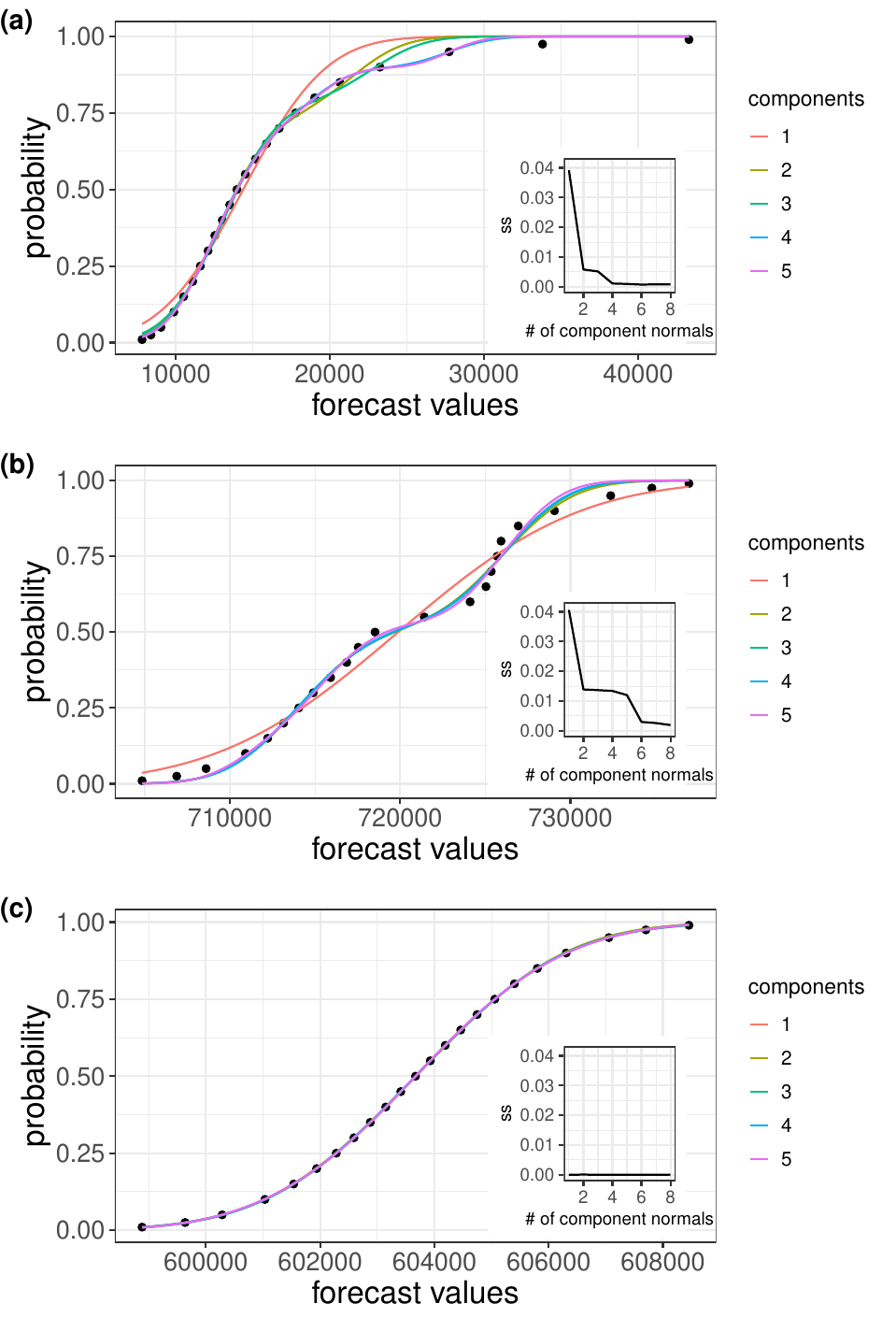}}
\begin{center}
\begin{minipage}{10cm}
\caption[Quantile forecasts with CDFs of fit mixture distributions]{Actual 
quantiles for three different COVID-19 forecasts with CDFs
of fit mixture distributions from one to five components. The inner plots 
show the SS value as the number of components is increased.}
\label{fig:quantfit1}
\end{minipage}
\end{center}
\end{figure}

Table \ref{table:maxss} shows the maximum SS values over all fits for fits from
one to five component mixture distributions. Figure \ref{fig:quantfit1} contains 
plots showing fits for selected individual forecasts with fit CDF functions
plotted over quantiles in the left plots, and SSE values plotted by the number
of components. 

\begin{table}[h!]
\begin{center}
\begin{minipage}{10cm}
\centering
\caption[Maximum SS between selected quantile forecasts and fit mixture
  distribution CDFs]{Maximum SS value over all quantile forecasts to which mixture
  distributions were fit for one to five components}
 \begin{tabular}{c|c c c c c}
 Components  & 1     & 2     & 3     & 4     & 5     \\
 \hline
   max SS   & 0.53  & 0.36  & 0.16  & 0.11  & 0.06   \\
\end{tabular}
  
\label{table:maxss}
\end{minipage}
\end{center}
\end{table}

\begin{figure}[htbp]
\centerline{\includegraphics[scale=.84]{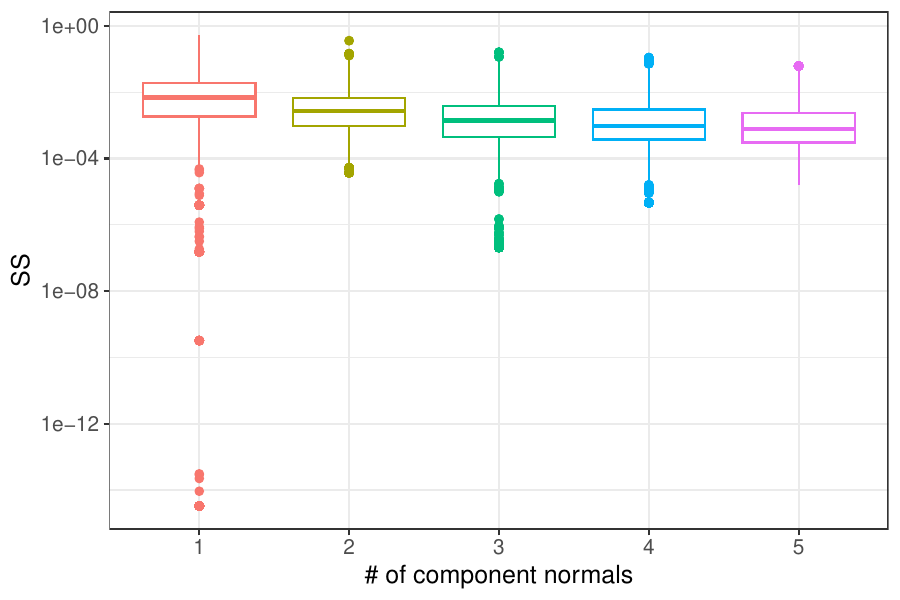}}
\begin{center}
\begin{minipage}{10cm}
\caption[Boxplots of SS (on log scale) between submitted forecasts and fit 
CDFs with increasing components in mixture distributions]{Boxplots of SS 
(on the log scale) between submitted quantile 
forecasts and fit CDFs}
\label{fig:ssbox}
\end{minipage}
\end{center}
\end{figure}

\begin{figure}[htbp]
\centerline{\includegraphics[scale=.8]{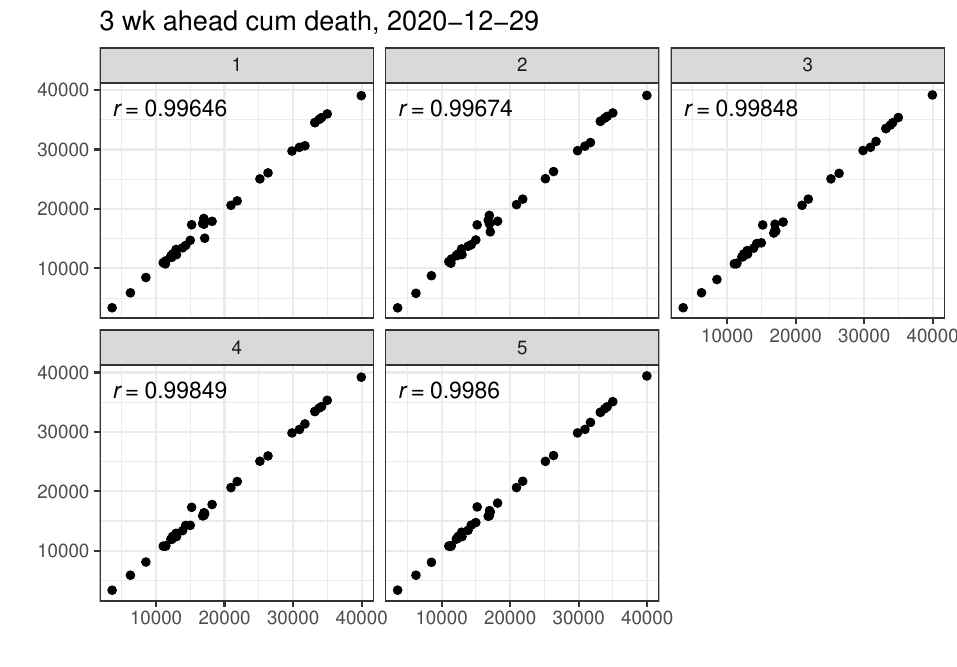}}
\begin{center}
\begin{minipage}{10cm}
\caption[Scatterplots for true WISs and WISs from fit mixture distributions]{
Scatter plots for actual WISs and WISs from fit 
mixture distributions for 3 week ahead COVID-19 forecasts from 12/29/2020}
\label{fig:wisscat}
\end{minipage}
\end{center}
\end{figure}

To compare forecast performance between actual quantile forecasts and the fit
distributions, truth for each week/target combination was obtained and the 
WIS from \eqref{eq:wis} was calculated. Figure \ref{fig:wisscat} shows plots 
of the WIS scores for 3 week ahead cumulative death forecasts for December 29,
2020 plotted against WIS scores for mixture distributions fit to the forecasts.
There is a separate scatterplot for fits of one to five components. Table 
\ref{table:wiscor} shows calculated correlations between actual WIS scores and 
fit WIS scores for 5  different week/target combinations.

\begin{table}[h!]
\begin{center}
\begin{minipage}{10cm}
\centering
\caption[Table of WIS correlations]{Correlation between actual WIS and WIS 
  from fit mixture 
  distributions for selected forecasts of the following dates and targets
  (a) 9/20/22 2 wk cum, (b) 9/14/21 1 wk inc, (c) 12/22/20 4 wk cum, 
  (d) 7/14/20 4 wk cum, (e) 12/22/20 2 wk cum}
 \begin{tabular}{c|c c c c c c} 
 Components  & Overall  & (a)      & (b)      & (c)      & (d)      & (e)     \\
 \hline
   1         & 0.99966  & 0.99999  & 0.99890  & 0.99851  & 0.99727  & 0.9968  \\
   2         & 0.99949  & 1        & 0.99913  & 0.99899  & 0.99762  & 0.99673 \\
   3         & 0.99959  & 1        & 0.99947  & 0.99961  & 0.99603  & 0.99819 \\
   4         & 0.9996   & 1        & 0.99945  & 0.99959  & 0.99808  & 0.99535 \\
   5         & 0.99978  & 1        & 0.99952  & 0.99962  & 0.99865  & 0.99788 \\
 \hline

 \end{tabular}
  
\label{table:wiscor}
\end{minipage}
\end{center}
\end{table}

The results from these studies suggest that forecasts in bin distribution and 
quantile formats may indeed be increasingly well approximated by a mixture of 
normal distributions as the number of components increases.

\subsection{Sample Distribution Forecast}
\label{section:sampdist}

As some forecasting projects may accept sample distributions for 
forecasts, we created this example to show that a sample may be closely 
approximated by a mixture distribution as the number of components increases.
To show this, we first created a sample distribution. With the sample in hand 
we fit a normal distribution by calclating the maximum likelihood estimates for
mean and standard deviation. We then fit mixture distributions from two to five
components to the sample, thus giving fits for mixture distributions of one to 
five component normal mixture distributions.

We obtained a sample distribution by first selecting a bin forecast from the flu
forecasting competition. We randomly drew 700 samples where each draw 
corresponded to a bin $B_i$ with probability according to the forecast $p_i$. 
For each draw corresponding to each bin $B_i$, a value was randomly selected 
according to the uniform distribition $Unif(b_{i-1},b_i)$. 

\begin{figure}[htbp]
\centerline{\includegraphics[scale=.89]{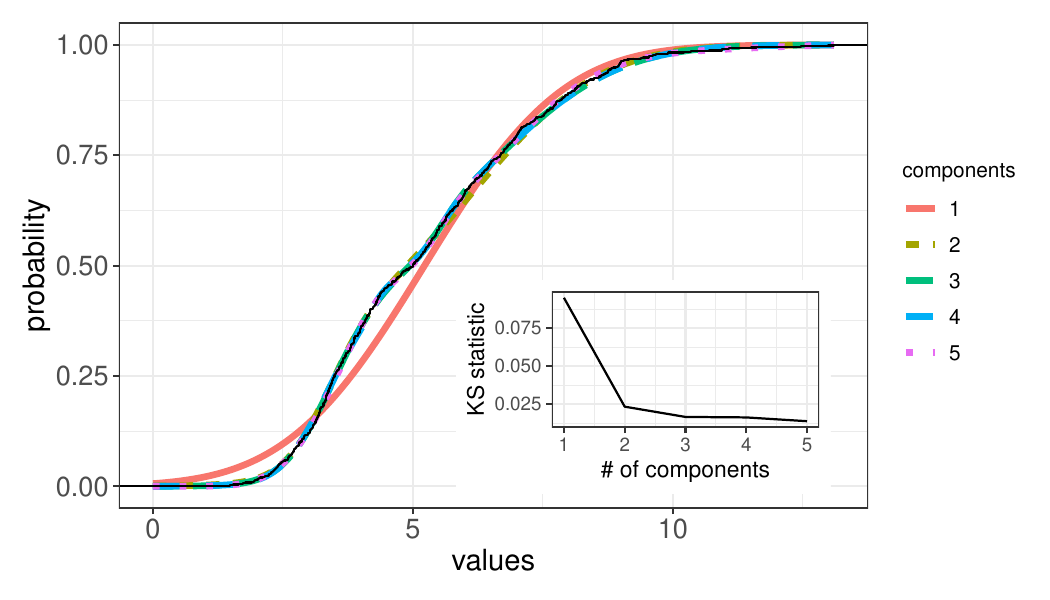}}
\begin{center}
\begin{minipage}{10cm}
\caption[Fit CDF and ECDF curves for example sample]{The outer plot shows the 
fit CDFs in color plotted with the ECDF in 
black of the sample. The inner plot shows the value of the KS statistic of the 
fit CDFs and the ECDF as the number of components increases.}
\label{fig:emecdf}
\end{minipage}
\end{center}
\end{figure}

\begin{figure}[htbp]
\centerline{\includegraphics[scale=.89]{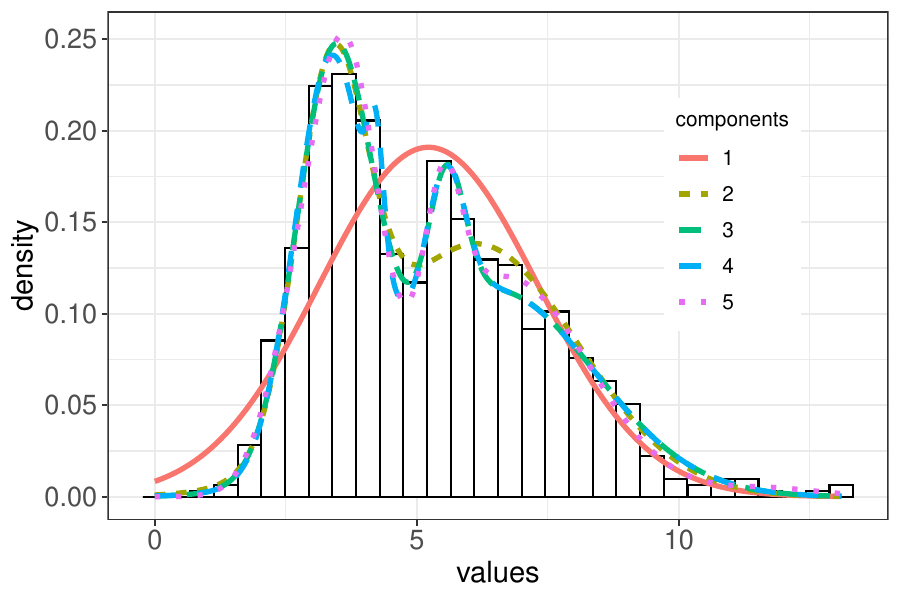}}
\begin{center}
\begin{minipage}{10cm}
\caption[Sample histogram with fit pdfs]{Histogram of the sample with pdfs of 
the fit mixture distributions}
\label{fig:emhist}
\end{minipage}
\end{center}
\end{figure}

The expected maximization (EM) algorithm was used to find maximum likelihood 
values to fit a mixture distribution to the sample. We used the function
\texttt{normalmixEM} in the \texttt{mixtools} package in \texttt{R} for this.
For each of the five fits, we calculated the Kolmogorov-Smirnov (KS) test 
statistic or the maximum distance between the ECDF function of the sample
and the CDF function of the fit mixture distrubtion. Figures \ref{fig:emecdf} 
and \ref{fig:emhist} show the fit CDFs plotted with the ECDF and the pdfs
plotted on the histogram respectively. These figures show that the sample is 
more closely approximated as the number of components in a mixture distribution 
increases.

\section{Discussion}

In this paper we have reviewed four representations commonly used in
probabilistic 
forecasting and discussed proper scoring, data storage, and ensemble model 
construction for each type. We presented the mixture distribution
representation and argue that its use in collaborative probabilistic 
forecasting is preferable to the other representations.
In terms of model flexibility, storage, and 
ensemble construction it is
comparable to bin and quantile forecasts but also provides a 
forecast with a infinite nominal resolution. Based on a retrospective analysis,
we argue that forecasts of quantile or bin distribution representations
may be more closely approximated by
a mixture distribution as the number of components of the distribution is 
increased. This may allow the 
transition from past and current formats to a mixture distribution format 
straightforward.
We thus advocate for the use of mixture distributions in
future forecasting projects like those done in the CDC flu competition or in the
COVID-19 Forecast Hub.

For a number of reasons, some forecasters may prefer not to the adopt mixture 
distributions as a format in collaborative forecasting. 
A collaborative forecast center, along with 
forecasters, using a 
different representation format may simply not want to break from tradition. 
There may be some concern that a mixture distribution does not represent well
certain models. And the 
implementation of new scoring and ensemble construction methods may also be a 
barrier. Development of tools beyond what was used in Section 
\ref{section:tools} would assist in making a transition to using mixture 
distributions more straightforward.
One aspect of ensemble construction which received little attention in 
this paper is the selection of weights for components of an ensemble where each
of the components is a mixture distribution. 
Computing requirements could be a
concern in such a problem, and further research on this may provide
ideas of best methods for weight selection.

Another area of recommended research is the use of joint mixture distributions 
for 
forecasting. We have only considered here probabilistic forecasting of one 
event at a time, or example, the number of new infections in one week at one 
specific location. 
This forecast is presented as a marginal distribution for that 
specific target, time, and location. A joint distribution for forecasting 
multiple targets, times, or locations may sometimes be desirable and may require
further consideration on how joint mixture distributions could be used as a 
format in collaboration.


\bibliographystyle{unsrt}  
\bibliography{references}


\newpage
\section{APPENDIX}
\label{section:append}

The following is the code to make the \texttt{MakeDist()} function introduced 
in \ref{section:tools}. Tables \ref{table:mixfams} and \ref{table:mixfamsparams}
show how distribution family arguments are to be written and what parameters are
to be used for the distributions in the \texttt{MakeDist()} function.

\begin{knitrout}
\definecolor{shadecolor}{rgb}{0.969, 0.969, 0.969}\color{fgcolor}\begin{kframe}
\begin{alltt}
\hlstd{MakeDist} \hlkwb{<-} \hlkwa{function}\hlstd{(}\hlkwc{distsdf}\hlstd{)\{}

  \hlstd{distdf} \hlkwb{<-} \hlstd{distsdf[distsdf[,}\hlnum{1}\hlstd{]} \hlopt{!=} \hlstr{'Lst'}\hlstd{,]}
  \hlstd{tdist} \hlkwb{<-} \hlstd{distsdf[distsdf[,}\hlnum{1}\hlstd{]} \hlopt{==} \hlstr{'Lst'}\hlstd{,]}

  \hlstd{fun_dist} \hlkwb{<-}
    \hlkwd{apply}\hlstd{(distdf,} \hlkwc{FUN}\hlstd{=}\hlkwa{function}\hlstd{(}\hlkwc{x}\hlstd{) \{}
      \hlkwd{paste}\hlstd{(}\hlstr{'distr::'}\hlstd{,x[}\hlnum{1}\hlstd{],} \hlstr{'('}\hlstd{,}
            \hlkwd{ifelse}\hlstd{(}\hlopt{!}\hlkwd{is.na}\hlstd{(x[}\hlnum{2}\hlstd{])} \hlopt{&} \hlstd{(}\hlopt{!}\hlkwd{is.na}\hlstd{(x[}\hlnum{3}\hlstd{])} \hlopt{| !}\hlkwd{is.na}\hlstd{(x[}\hlnum{4}\hlstd{])),}
                   \hlkwd{paste}\hlstd{(x[}\hlnum{2}\hlstd{],}\hlstr{','}\hlstd{,}\hlkwc{sep}\hlstd{=}\hlstr{''}\hlstd{),}
                   \hlkwd{ifelse}\hlstd{(}\hlopt{!}\hlkwd{is.na}\hlstd{(x[}\hlnum{2}\hlstd{])} \hlopt{&} \hlkwd{is.na}\hlstd{(x[}\hlnum{3}\hlstd{])} \hlopt{&} \hlkwd{is.na}\hlstd{(x[}\hlnum{4}\hlstd{]),}
                          \hlstd{x[}\hlnum{2}\hlstd{],} \hlstr{''}\hlstd{)),}
            \hlkwd{ifelse}\hlstd{(}\hlopt{!}\hlkwd{is.na}\hlstd{(x[}\hlnum{3}\hlstd{])} \hlopt{& !}\hlkwd{is.na}\hlstd{(x[}\hlnum{4}\hlstd{]),}
                   \hlkwd{paste}\hlstd{(x[}\hlnum{3}\hlstd{],}\hlstr{','}\hlstd{,}\hlkwc{sep}\hlstd{=}\hlstr{''}\hlstd{),}
                   \hlkwd{ifelse}\hlstd{(}\hlopt{!}\hlkwd{is.na}\hlstd{(x[}\hlnum{3}\hlstd{])} \hlopt{&} \hlkwd{is.na}\hlstd{(x[}\hlnum{4}\hlstd{]), x[}\hlnum{3}\hlstd{],}\hlstr{''}\hlstd{)),}
            \hlkwd{ifelse}\hlstd{(}\hlopt{!}\hlkwd{is.na}\hlstd{(x[}\hlnum{4}\hlstd{]),x[}\hlnum{4}\hlstd{],}\hlstr{''}\hlstd{),} \hlstr{')'}\hlstd{,}\hlkwc{sep}\hlstd{=}\hlstr{''}\hlstd{)}
    \hlstd{\},} \hlkwc{MARGIN} \hlstd{=} \hlnum{1}
    \hlstd{)}

  \hlstd{fun_tdist} \hlkwb{<-} \hlkwd{apply}\hlstd{(tdist,} \hlkwc{FUN}\hlstd{=}\hlkwa{function}\hlstd{(}\hlkwc{x}\hlstd{) \{}
    \hlkwd{paste0}\hlstd{(}\hlstr{'distr::Td('}\hlstd{,x[}\hlnum{4}\hlstd{],}\hlstr{')*'}\hlstd{,x[}\hlnum{3}\hlstd{],} \hlstr{'+'}\hlstd{, x[}\hlnum{2}\hlstd{])}
  \hlstd{\},} \hlkwc{MARGIN} \hlstd{=} \hlnum{1}
  \hlstd{)}

  \hlstd{dist_args} \hlkwb{<-} \hlkwd{paste}\hlstd{(fun_dist,} \hlkwc{collapse}\hlstd{=}\hlstr{','}\hlstd{,}\hlkwc{sep}\hlstd{=}\hlstr{''}\hlstd{)}
  \hlstd{tdist_args} \hlkwb{<-} \hlkwd{paste0}\hlstd{(fun_tdist,}\hlkwc{collapse}\hlstd{=}\hlstr{','}\hlstd{)}
  \hlstd{args} \hlkwb{<-} \hlkwd{ifelse}\hlstd{(tdist_args}\hlopt{!=}\hlstr{''}\hlstd{,}\hlkwd{paste}\hlstd{(dist_args,tdist_args,}\hlkwc{sep}\hlstd{=}\hlstr{','}\hlstd{),dist_args)}

  \hlstd{weights} \hlkwb{<-} \hlkwd{c}\hlstd{(distdf[,}\hlnum{5}\hlstd{],tdist[,}\hlnum{5}\hlstd{])}
  \hlstd{mixString} \hlkwb{<-} \hlkwd{paste}\hlstd{(}\hlstr{'distr::UnivarMixingDistribution('}\hlstd{,}
                     \hlstd{args,}\hlstr{',mixCoeff=weights)'}\hlstd{,}\hlkwc{sep}\hlstd{=}\hlstr{''}\hlstd{)}
  \hlstd{mixDist} \hlkwb{<-} \hlkwd{eval}\hlstd{(}\hlkwd{parse}\hlstd{(}\hlkwc{text}\hlstd{=mixString))}

  \hlkwd{return}\hlstd{(mixDist)}
\hlstd{\}}
\end{alltt}
\end{kframe}
\end{knitrout}

\begin{table}
\begin{center}
\caption[Arugments of \texttt{MakeDist()} function]{This table describes how
  each argument may be entered into a data frame entered into the 
  \texttt{MakeDist()} function, including
  exactly how the distribution families should appear and what paramaters should
  be included for each distribution family.}
    \begin{tabular}{ | l | p{4cm} | p{7cm} |}
    \hline
    \textbf{Argument} & \textbf{Summary} & \textbf{Options} \\ \hline
    dist & A string specifying a distribution family.  & Beta, Cauchy,
    Lnorm, Logis, “Unif, Lst (location scale t distribution),
    Weibull, Fd, Norm,  Chisq,
    Gammad, Exp
	 Binom, Dirac, Pois, Hyper, Nbinom, Geom \\ \hline
    param1 & A real number specifying the first parameter value of the
    distribution. & Beta: shape1;
	Cauchy: location;
	Lnorm: meanlog;
	Logis: location;
	Unif: min;
	Lst: location;
	Weibull: shape;
	Fd: df1;
	Norm: mean;
	Chisq: df;
	Gammad: scale;
	Exp: rate;
	Binom: size;
	Dirac: location;
	Pois: lambda;
	Hyper: m;
	Nbinom: n;
	Geom: prob
     \\ \hline
    param2 & A real number specifying the second parameter value of the
    distribution. & Beta: shape2;
	Cauchy: scale;
	Lnorm: slog;
	Logis: scale;
	Unif: Max;
	Lst: scale;
	Weibull: scale;
	Fd: df2;
	Norm: sd;
	Chisq: ncp;
	Gammad: shape;
	Binom: prob;
	Hyper: n;
	Nbinom: p \\
    \hline
    param3 & A real number specifying the third parameter value of the
    distribution. & Lst: df;
	Hyper: k \\
	\hline
	weight & A real number between 0 and 1 specifying the weight given to the
	distribution in the overall mixture distribution. The sum of the weight
	column should equal 1. & ...
	\\
	\hline
    \end{tabular}
    \begin{center}
\begin{minipage}{10cm}
  
  \label{table:mixfams}
  \end{minipage}
  \end{center}
\end{center}
\end{table}

\begin{table}
\begin{center}
\caption[\texttt{MakeDist()} parameters]{This table shows what parameters
  may be used for each distribution family used in a data frame for
  the \texttt{MakeDist()} function or in the 
  \texttt{UnivarMixingDistribution()} function.}
    \begin{tabular}{ l l  p{2cm}  p{2cm}  p{2cm}}
    \hline
    \textbf{Distribution} & \textbf{dist} & \textbf{param1} & \textbf{param2} &
    \textbf{param3} \newline \\ \hline
    Beta & Beta & shape1 & shape2 &  \\
    Cauch & Cauchy & location & scale &  \\
    Log-normal & Lnorm & meanlog & slog &  \\
    Logistic & Logis & location & scale &  \\
    Uniform & Unif & min & max &  \\
    Location Scale T & Lst & location & scale & df \\
    Weibull & Weibull & shape & scale &  \\
    F & Fd & df1 & df2 &  \\
    Normal & Norm & mean & sd &  \\
    Chisqure & Chisq & df &  &  \\
    Gamma & Gammad & scale & shape &  \\
    Exponential & Exp & rate &  &  \\
    Binomial & Binom & size & prob &  \\
    Dirac & Dirac & location &  &  \\
    Poisson & Pois & lambda &  &  \\
    Hypergeometric & Hyper & m & n & k \\
    Negative binomial & Nbinom & n & p &  \\
    Geometric & Geom & prob &  &  \\ \hline
    \end{tabular}
        \begin{center}
\begin{minipage}{10cm}
  
  \label{table:mixfamsparams}
  \end{minipage}
  \end{center}
\end{center}
\end{table}

The following is the code used to make the function \texttt{CRPS()} used in
section \ref{section:tools}.
\begin{knitrout}
\definecolor{shadecolor}{rgb}{0.969, 0.969, 0.969}\color{fgcolor}\begin{kframe}
\begin{alltt}
\hlstd{crps_integrand} \hlkwb{<-} \hlkwa{function}\hlstd{(}\hlkwc{x}\hlstd{,}\hlkwc{dist}\hlstd{,}\hlkwc{y}\hlstd{) \{(}\hlkwd{dist}\hlstd{(x)} \hlopt{-} \hlkwd{as.numeric}\hlstd{(y} \hlopt{<=} \hlstd{x))}\hlopt{^}\hlnum{2}\hlstd{\}}

\hlstd{CRPS} \hlkwb{<-} \hlkwa{function}\hlstd{(}\hlkwc{y}\hlstd{,}\hlkwc{dist}\hlstd{) \{}
  \hlstd{int} \hlkwb{<-} \hlkwd{integrate}\hlstd{(crps_integrand,}\hlopt{-}\hlnum{Inf}\hlstd{,}\hlnum{Inf}\hlstd{,y,}\hlkwc{dist}\hlstd{=dist)}
  \hlkwd{return}\hlstd{(int}\hlopt{$}\hlstd{value)}
\hlstd{\}}
\end{alltt}
\end{kframe}
\end{knitrout}



\end{document}